%% file: main.tex
\documentclass[conference]{IEEEtran}

\usepackage{amsmath}
\usepackage{amsfonts}
\usepackage{amsthm}
\usepackage{stmaryrd}
\usepackage{amssymb}
\usepackage{xspace}
\usepackage{verbatim}
\usepackage{graphicx}
\usepackage{array}
\usepackage{tikz}
\usetikzlibrary{arrows,positioning}

\input{macro}

\IEEEoverridecommandlockouts

\begin{document}

\title{From Two-\!Way to One-\!Way Finite State Transducers~\thanks{This work has been partly supported by the project ECSPER funded by the french agency for research (ANR-09-JCJC-0069), by the project SOSP funded by the CNRS, and by the Faculty of Sciences of University Paris-Est Cr\'eteil.}}
\author{\IEEEauthorblockN{Emmanuel Filiot}
  \IEEEauthorblockA{LACL}
  \IEEEauthorblockA{University Paris-Est Cr\'eteil}
  \and
  \IEEEauthorblockN{Olivier Gauwin}
  \IEEEauthorblockA{LaBRI,}
  \IEEEauthorblockA{University of Bordeaux}
  \and
  \IEEEauthorblockN{Pierre-Alain Reynier}
  \IEEEauthorblockA{LIF, Aix-Marseille Univ.}
  \IEEEauthorblockA{\& CNRS, UMR 7279}
  \and
  \IEEEauthorblockN{Fr\'ed\'eric Servais}
  \IEEEauthorblockA{Hasselt University and}
   \IEEEauthorblockA{Transnational University of Limburg}
}

\maketitle

\begin{abstract}
    Any two-way finite state automaton is equivalent to some one-way finite state automaton. 
    This well-known result, shown by Rabin and Scott and independently 
    by Shepherdson, states that two-way finite state automata (even non-deterministic) characterize
    the class of regular languages. 
    It is also known that this result does not extend to finite string transductions:
    (deterministic) two-way finite state transducers strictly extend the expressive power of
    (functional) one-way transducers. In particular deterministic two-way transducers capture exactly the class
    of MSO-transductions of finite strings.

    In this paper, we address the following definability problem: given 
    a function defined by a two-way finite state transducer, is it definable by 
    a one-way finite state transducer? By extending Rabin and Scott's proof to 
    transductions, we show that this problem is decidable.
    Our procedure builds a one-way transducer, which is equivalent to 
    the two-way transducer, whenever one exists.
    
\end{abstract}

 


\input{introduction}
\input{prelim}

\input{generalcase}
\input{zmotions}

\input{complexity}

\input{application}

\bibliographystyle{IEEEtran}
\bibliography{papers}

\newpage
\appendices
\input{appendix}

\end{document}

%% file: macro.tex
\newcommand{\olivier}[1]{\quad\\\fbox{\parbox{\linewidth}{{\sc Olivier}:\\#1}}}

\newcommand\openquestion[1]%
  {\quad\\\fbox{\parbox{\linewidth}{{\sc Open Question}:\\#1}}}
\newcommand{\ignore}[1]{}

\newcommand{\lang}[1]{\mathcal{L}(#1)}
\newcommand{\primroot}{\ensuremath{\mu}}

\newcommand{\shape}{\mathrm{shape}}

\newcommand{\Zmode}{\textit{Z}-mode\xspace}
\newcommand{\Tmode}{\textit{T}-mode\xspace}
\newcommand{\acrostyle}[1]{\textit{#1}\xspace}
\newcommand{\ZNFT}{\acrostyle{ZNFT}}
\newcommand{\ZNFTs}{\acrostyle{ZNFTs}}
\newcommand{\eZNFT}{\acrostyle{$\epsilon$ZNFT}}
\newcommand{\eZNFTs}{\acrostyle{$\epsilon$ZNFTs}}

\newcommand{\NFT}{\acrostyle{NFT}}
\newcommand{\NFTs}{\acrostyle{NFTs}}
\newcommand{\DFT}{\acrostyle{DFT}}
\newcommand{\DFTs}{\acrostyle{DFTs}}
\newcommand{\FUNTW}{\acrostyle{f2NFT}}
\newcommand{\FUNTWs}{\acrostyle{f2NFTs}}
\newcommand{\FUNNFT}{\acrostyle{fNFT}}

\newcommand{\TWNFT}{\acrostyle{2NFT}}
\newcommand{\TWNFTs}{\acrostyle{2NFTs}}
\newcommand{\TWDFT}{\acrostyle{2DFT}}
\newcommand{\TWDFTs}{\acrostyle{2DFTs}}
\newcommand{\NFA}{\acrostyle{NFA}}
\newcommand{\NFAs}{\acrostyle{NFAs}}
\newcommand{\TWNFA}{\acrostyle{2NFA}}
\newcommand{\TWNFAs}{\acrostyle{2NFAs}}
\newcommand{\DFA}{\acrostyle{DFA}}
\newcommand{\DFAs}{\acrostyle{DFAs}}

\newcommand{\TWDFAs}{\acrostyle{2DFAs}}
\newcommand{\out}{\mathrm{out}}
\newcommand{\zmotion}{$z$-motion\xspace}
\newcommand{\zmotions}{$z$-motions\xspace}

\newcommand{\lc}{\mathcal{L}}
\newcommand{\rc}{\mathcal{R}}

\newcommand{\ppt}{\mathcal{P}}
\newcommand{\pptone}{\mathcal{P}_1}
\newcommand{\ppttwo}{\mathcal{P}_2}

\newcommand{\reversal}[2]{{\mathsf{rev}_{#1}(#2)}}

\newcommand{\intvSet}[1]{\mathcal{I}(#1)}
\newcommand{\intvForest}[1]{\mathcal{F}(#1)}
\newcommand{\squeezeOp}{\mathrm{squeeze}\xspace}
\newcommand{\squeeze}[1]{\squeezeOp(#1)}
\newcommand{\squeezepow}[2]{\squeezeOp^{#1}(#2)}
\newcommand{\sq}[1]{\mathrm{sq}(#1)}
\newcommand{\sqn}[2]{\mathrm{sq}^{#1}(#2)}
\newcommand{\revpos}[3]{\mathsf{revpos}_{#2}^{#1}(#3)}
\newcommand{\nested}[2]{(#1)<_\mathrm{nest}(#2)}
\newcommand{\nestlevel}[1]{\mathrm{nestLevel}(#1)}

\newcommand{\dom}{\mathrm{dom}}

\newtheorem{proposition}{Proposition}
\newtheorem{definition}{Definition}

\newtheorem{lemma}{Lemma}
\newtheorem{theorem}{Theorem}

%% file: introduction.tex
\section{Introduction}


In formal language theory, the importance of a class of languages is often supported by
the number and the diversity of its characterizations. One of the most famous  
example is the class of regular languages of finite strings,
which enjoys, for instance, computational (automata), algebraic (syntactic congruence) and logical (monadic second order (MSO) logic with one successor) characterizations. The study of regular languages has been very influential and several 
generalizations have been established. Among the most notable ones are the extensions to infinite strings \cite{Buchi:62} and 
trees \cite{Thatcher:Wright:mst:1968}. On finite strings, it is well-known that both deterministic and non-deterministic
finite state automata define regular languages. It is also well-known that the expressive power of finite state automata
does not increase when the reading head can move left and right, even in presence of non-determinism. The latter class
is known as non-deterministic \emph{two-way finite state automata} and it is no more powerful than
(one-way) finite state automata. The proof of this result was first shown in the seminal paper of Rabin and Scott \cite{RabinScott59}, and 
 independently by Shepherdson \cite{Shepherdson59}.

The picture of automata models over finite strings changes substantially when,
instead of languages,  string \emph{transductions}, i.e. relations from strings to strings,  are considered. \emph{Transducers}
generalize automata as they are equipped with a one-way output tape. At each step they 
read an input symbol, they can append several symbols to the output tape. Their transition
systems can be either deterministic or non-deterministic. \emph{Functional} transducers are 
transducers that define functions instead of relations. For instance, deterministic transducers
are always functional.  In this paper, 
we are interested in transducers that define functions, but that can be non-deterministic.

As for automata, the reading head of transducers can move one-way (left-to-right) or two-way.
 \emph{(One-way) finite state  transducers} have been extensively studied \cite{Berstel79,sakarovich:2009a}.  Non-deterministic (even functional) one-way transducers (\NFTs) strictly extend the expressive power of deterministic one-way transducers (\DFTs), because non-determinism allows one to express local transformations
that depend on properties of the future of the input string.

 \emph{Two-way finite state transducers} define regular transformations that are beyond 
the expressive power of one-way transducers \cite{Cou-expression-gpg}. 
They can for instance reverse an input string, swap two substrings or copy a substring.
The transductions defined by two-way transducers have been characterized by other
logical and computational models. Introduced by Courcelle, monadic second-order definable transductions
are transformations from graphs to graphs defined with the logic MSO \cite{Cou94}. Engelfriet and Hoogeboom
have shown that the monadic second-order definable functions are exactly the functions definable
by deterministic two-way finite state transducers (\TWDFTs) when the graphs are restricted to finite strings \cite{EngHoo01}. Recently, Alur and {\v C}ern\'y have characterized \TWDFT-definable transductions by
a deterministic one-way model called \emph{streaming string transducers} \cite{alur_et_al:LIPIcs:2010:2853} and shown how they can be applied to the verification of list-processing programs \cite{conf/popl/AlurC11}. Streaming string transducers extend \DFTs with a finite set of output string variables. At each step, their content
can be reset or updated by either prepending or appending a finite string, or the content
of another variable, in a copyless manner.  Extending \TWDFTs with non-determinism
does not increase their expressive power when they define functions: 
non-deterministic two-way finite state transducers (\TWNFTs) that are functional define exactly
the class of functions definable by \TWDFTs \cite{EngHoo01,DBLP:conf/lata/Souza13}.
To summarize, there is a strict hierarchy between \DFT-, functional \NFT- and \TWDFT-definable
transductions.

%
%

Several important problems are known to be decidable for one-way transducers.
The \emph{functionality} problem for \NFT, 
decidable in PTime \cite{GurIba83,BealEtAl03a}, asks whether a given \NFT is functional. The \emph{determinizability} problem, also
decidable in PTime \cite{DBLP:journals/iandc/WeberK95,BealEtAl03a},  asks whether a given functional \NFT can be determinized, i.e. defines a \emph{subsequential} function. Subsequential functions are those functions that can be defined 
by \DFTs equipped with an additional output function from final states to finite strings, 
which is used to append a last string to the output when the computation terminates
successfully in some final state. Over strings that always end with a unique end marker, subsequential functions  are exactly the functions definable by \DFTs. 
For \TWNFTs, the functionality problem is known to be decidable  \cite{CulKar87}. Therefore the determinizability problem is also decidable for \TWNFTs, since
functional \TWNFTs and \TWDFTs have the same expressive power.
In the same line of research, we address a definability problem in this paper. 
In particular we answer the fundamental question of \NFT-definability of transductions 
defined by functional \TWNFTs.

\begin{theorem}\label{thm:main}
For all functional \TWNFTs $T$, it is decidable whether 
the transduction defined by $T$ is definable by an \NFT.
\end{theorem}

The proof of Theorem \ref{thm:main} extends the proof of Rabin and
Scott \cite{RabinScott59} from automata to
transducers\footnote{Shepherdson \cite{Shepherdson59} and then Vardi
  \cite{Vardi89} proposed arguably simpler constructions for automata.
  It is however not clear to us how to extend these constructions to
  transducers.}.  The original proof of Rabin and Scott is based on
the following observation about the runs of two-way automata. Their
shapes have a nesting structure: they are composed of many zigzags,
each zigzag being itself composed of simpler zigzags. Basic zigzags
are called \zmotions as their shapes look like a $Z$. Rabin and Scott
prove that for automata, it is always possible to replace a \zmotion
by a single pass.
Then from a two-way automaton $A$ it is possible to construct an
equivalent two-way automaton $B$ (called the squeeze of $A$) which is
simpler in the following sense: accepting runs of $B$ are those of $A$
in which some \zmotions have been replaced by single pass runs.
Last, they argue\footnote{To our knowledge, there is no published
  proof of this result, thus we prove it in this paper as we use it
  for transducers.} that after a number of applications of
this construction that depends only on the number of states of $A$,
every zigzag can be removed, yielding an equivalent one-way automaton. 

The extension to \TWNFTs faces the following additional difficulty:
it is not always possible to replace a \zmotion of a transducer by
a single pass. Intuitively, this is due to the fact that \TWNFTs
are strictly more expressive than \NFTs.
As our aim is to decide when a \TWNFT $T$ is \NFT-definable, we need
to prove that the \NFT-definability of $T$ implies that of every
\zmotion of $T$, to be able to apply the squeeze construction. The
main technical contribution of this paper is thus the study of the
\NFT-definability of \zmotions of transducers.  We show that this
problem is decidable, and identify a characterization which allows one to
prove that the \NFT-definability of $T$ implies that of every
\zmotion of $T$.

This characterization expresses requirements about the output strings
produced along loops of \zmotions. 
We show that when \zmotions 
are \NFT-definable, the output strings produced by the three passes on
a loop are not arbitrary, but conjugates. This allows us to give a
precise characterization of the form of these output strings.
We show that it is decidable to check whether all outputs words have
this form. Last, we present how to use this characterization to
simulate an \NFT-definable \zmotion by a single pass.

\vspace{1mm}
\noindent\textbf{Applications} By Theorem \ref{thm:main} and since functionality is decidable for \TWNFTs, it is also decidable, 
given  a \TWNFT, whether the transduction it defines is definable
by a functional \NFT. Another corollary of Theorem \ref{thm:main} and
the fact that functionality of \TWNFTs and
determinizability of \NFTs are both decidable is the  following theorem:
\begin{theorem}\label{thm:main2}
    For all \TWNFTs $T$, it is decidable whether 
    the transduction defined by $T$ is a subsequential function.
\end{theorem}

A practical application of this result lies in the static analysis of
memory requirements for evaluating (textual and functional) document transformations
in a streaming fashion. In this scenario, the input string is received as a
left-to-right stream. When the input stream is huge, it should not be entirely
loaded in memory but rather processed on-the-fly. Similarly, the output
string should not be stored in memory but produced as a stream. 
The remaining amount of memory needed to evaluate the transformation characterizes
its streaming space complexity. \emph{Streamable} transformations are those
transformations for which the required memory is bounded by a constant, 
and therefore is independent on the length of the input stream. It is known that 
streamable transformations correspond to transformations definable by subsequential
(functional) \NFTs \cite{conf/fsttcs/FiliotGRS11}. The \emph{streamabability} problem
asks, given a transformation defined by some transducer, whether it is streamable. 
Therefore for transformations defined by functional \NFTs, streamability coincides
with determinizability, and is decidable in PTime \cite{DBLP:journals/iandc/WeberK95,BealEtAl03a}.
Theorem \ref{thm:main2} is a generalization of this latter result to regular transformations, 
i.e. transformations defined by functional \TWNFTs, MSO transducers or streaming string transducers \cite{alur_et_al:LIPIcs:2010:2853}.
Other streamability problems have been studied for XML validation \cite{conf/icdt/SegoufinS07,conf/stacs/BaranyLS06}, 
XML queries \cite{journals/iandc/GauwinNT11} and XML transformations  \cite{conf/fsttcs/FiliotGRS11}. However the XML tree
transformations of \cite{conf/fsttcs/FiliotGRS11} are incomparable with the regular string transformations studied
in this paper.

\vspace{1mm}
\noindent \textbf{Related work} Most of the related work has already been mentioned. 
To the best of our knowledge, it is the first result that addresses a definability problem
between two-way and one-way transducers. In \cite{conf/dlt/Carton12}, two-way transducers
with a two-way output tape are introduced with a special output policy: each time a cell
at position $i$ of the input tape is processed, the output is written in the cell at position $i$
of the output tape. With that restriction, it is shown that two-way and one-way
transducers (\NFTs) define the same class of functions. 
In \cite{Anselmo90}, the result of Rabin and Scott, and Shepherdson,
is extended to two-way automata with multiplicities. In this context,
two-way automata strictly extend one-way automata.

\vspace{1mm}
\noindent\textbf{Organization of the paper} Section~\ref{sec:prelim} introduces necessary preliminary definitions. In Section~\ref{sec:generalcase}, 
we describe the general decision procedure for testing \NFT-definability of functional \TWNFTs.
We introduce \zmotion transductions induced by \TWNFTs and show that
their \NFT-definability is necessary. The decidability of this necessary condition as well as the construction
from \zmotion transducers to \NFTs are the most technical results of this paper and are the subject of Section~\ref{sec:zmotions}. 
We finally discuss side results and 
further questions in Section~\ref{sec:discussion}.

%% file: prelim.tex
\section{One-Way and Two-Way Finite State Machines}\label{sec:prelim}

\noindent\textbf{Words, Languages and Transductions}
Given a finite alphabet $\Sigma$, we denote by
$\Sigma^*$ the set of finite words over $\Sigma$, and by
$\epsilon$ the empty word.
The length of a word $u\in\Sigma^*$ is its number of
symbols, denoted by $|u|$. For all $i\in\{1,\dots,|u|\}$, we
denote by $u[i]$ the $i$-th letter of $u$. Given $1\leq i\leq j\leq |u|$, we
denote by $u[i..j]$ the word $u[i]u[i+1]\dots u[j]$ and by
$u[j..i]$ the word $u[j]u[j-1]\dots u[i]$.
We say that $v\in\Sigma^*$ is a \emph{factor} of $u$ if there exist $u_1,u_2\in\Sigma^*$ such that
$u = u_1vu_2$.
By $\overline{u}$ we denote the \emph{mirror} of $u$, i.e. the word of length
$|u|$ such that $\overline{u}[i]=u[|u|-i+1]$ for all $1\le i\le |u|$.

The \emph{primitive root} of $u\in\Sigma^*$ is the shortest
word $v$ such that $u = v^k$ for some integer $k\geq 1$, and is
denoted by $\primroot(u)$. 
Two words $u$ and $v$ are \emph{conjugates}, denoted by 
$\sim$, if there exist
$x,y\in\Sigma^*$ such that $u = xy$ and $v = yx$, i.e.
$u$ can be obtained from $v$ by a cyclic permutation. Note that
$\sim$ is an equivalence relation.
We will use this fundamental lemma:
\begin{lemma}[\cite{Choffrut97}]\label{lem:fundamental}
  Let $u,v\in\Sigma^*$. If there exists $n\geq 0$ such that
  $u^n$ and $v^n$ have a common factor of length at least
  $|u|+|v|-gcd(|u|,|v|)$, then $\primroot(u)\sim \primroot(v)$.
\end{lemma}

Note that if $\primroot(u)\sim \primroot(v)$, then there exist $x,y\in\Sigma^*$ such that
$u \in (xy)^*$ and $v\in (yx)^*$.

A \emph{language} over $\Sigma$ is a set $L\subseteq \Sigma^*$. A \emph{transduction}
over $\Sigma$ is a relation $R \subseteq \Sigma^*\times \Sigma^*$. Its domain is denoted
by $\dom(R)$, i.e. $\dom(R) = \{ u\ |\ \exists v,\ (u,v)\in R\}$, while its image 
$\{ v\ |\ \exists u,\ (u,v)\in R\}$ is denoted by $img(R)$. A transduction $R$ is
\emph{functional} if it is a function.

\vspace{1mm}
\noindent\textbf{Automata}
A \emph{non-deterministic two-way finite state automaton}\footnote{We follow the definition of Vardi \cite{Vardi89}, but
without stay transitions. This is without loss of generality though.} (\TWNFA) over a finite alphabet
$\Sigma$ is a tuple $A = (Q, q_0, F, \Delta)$  where $Q$ is a finite set of states, $q_0\in Q$ is 
the initial state, $F\subseteq Q$ is a set of final states, and $\Delta$ is the
transition relation, of type $\Delta \subseteq Q\times \Sigma\times Q\times \{+1,-1\}$. 
It is \emph{deterministic} if for all $(p,a)\in Q\times \Sigma$, there is at most one pair
$(q,m)\in Q\times \{+1,-1\}$ such that $(p,a,q,m)\in \Delta$.
In order to see how  words are evaluated by $A$, it is convenient to see the input as 
a right-infinite input tape containing the word (starting at the first cell) followed by blank symbols.
Initially the head of $A$ is on the first cell in state $q_0$ (the cell at position $1$). When $A$ reads an input symbol, depending on the
transitions in $\Delta$, its head moves to the left ($-1$) if the head was not in the first cell, 
or to the right ($+1$) and changes its state. $A$ stops as soon as it reaches a blank symbol (therefore at the right of
the input word), and the word is accepted if the current state is final.

A \emph{configuration} of $A$ is a pair $(q,i)\in Q\times (\mathbb{N}-\{0\})$ where $q$ is a state and $i$ is a position on the input tape.
A \emph{run} $\rho$ of $A$ is a finite sequence of configurations. 
The run $\rho = (p_1,i_1)\dots (p_m,i_m)$ is a run 
on an input word $u\in\Sigma^*$ of length $n$ if $p_1 = q_0$, $i_1 = 1$, $i_m\leq n+1$, and for all $k\in \{1,\dots,m-1\}$, 
$1\leq i_k\leq n$ and $(p_k,u[i_k], p_{k+1}, i_{k+1}-i_k)\in \Delta$. It is \emph{accepting} if $i_m = n+1$ and $p_m \in F$.
The language of a \TWNFA $A$, denoted by $L(A)$, is the set of words $u$ such that there exists an accepting run
of $A$ on $u$.

A \emph{non-deterministic (one-way) finite state automaton} (\NFA) is a \TWNFA such that $\Delta\subseteq Q\times \Sigma \times Q \times \{+1\}$, therefore
we will often see $\Delta$ as a subset of $Q\times \Sigma \times Q$. 
Any \TWNFA is effectively equivalent to an \NFA. It was first proved by Rabin and Scott, and independently by Shepherdson \cite{RabinScott59,Shepherdson59}.



\noindent\textbf{Transducers}
\emph{Non-deterministic two-way finite state transducers} (\TWNFTs) over $\Sigma$ extend \NFAs with a one-way left-to-right output tape.
They are defined as \TWNFAs except that the transition relation $\Delta$ is extended with outputs:
$\Delta\subseteq Q\times \Sigma \times \Sigma^* \times Q \times \{-1,+1\}$. If a transition $(q,a,v,q',m)$ is fired
on a letter $a$, the word $v$ is appended to the right of the output tape and the transducer goes to state $q'$.
Wlog we assume that for all $p,q\in Q$, $a\in \Sigma$ and $m\in\{+1,-1\}$, there exists at most one $v\in\Sigma^*$
such that $(p,a,v,q,m)\in \Delta$. We also denote $v$ by $\out(p,a,q,m)$. 

A run of a \TWNFTs is a run of its underlying automaton, i.e. the \TWNFAs obtained by ignoring the output.
A run $\rho$ may be simultaneously a run on a word $u$ and on a word
$u'\neq u$. However, when the underlying input word is given, there is
a unique sequence of transitions associated with $\rho$. Given a
\TWNFT $T$, an input word $u\in\Sigma^*$ and a run $\rho =
(p_1,i_1)\dots (p_m,i_m)$ of $T$ on $u$, the output of $\rho$ on $u$, denoted by
$\out^u(\rho)$, is the word obtained by concatenating the outputs of
the transitions followed by $\rho$, i.e.  $\out^u(\rho) =
\out(p_1,u[i_1], p_2, i_2{-}i_1)\cdots \out(p_{m-1},u[i_{m-1}], p_m,
i_m{-}i_{m-1})$. If $\rho$ contains a single configuration, we let
$\out^u(\rho) = \epsilon$.  When the underlying input word $u$ is
clear from the context, we may omit the exponent $u$. The transduction
defined by $T$ is the relation $R(T) = \{(u,\out^u(\rho))\ |\
\rho\text{ is an accepting run of } T \text{ on } u\}$. We may often
just write $T$ when it is clear from the context. A \TWNFT $T$ is
\emph{functional} if the transduction it defines is functional. The
class of functional \TWNFTs is denoted by \FUNTW. In this paper, we
mainly focus on \FUNTWs.  The \emph{domain} of $T$ is defined as
$\dom(T) = \dom(R(T))$. The domain $\dom(T)$ is a regular language that
can be defined by the \TWNFA obtained by projecting away the output
part of the transitions of $T$, called the \emph{underlying input
  automaton}. A \emph{deterministic two-way finite state transducer}
(\TWDFT) is a \TWNFT whose underlying input automaton is
deterministic. Note that \TWDFTs are always functional, as there is at
most one accepting run per input word.
A \emph{non-deterministic (one-way) finite state transducer} (\NFT) is a \TWNFT whose underlying automaton is an \NFA\footnote{This definition implies that there is no $\epsilon$-transitions that can produce outputs, which may cause the image of an input word to be an infinite language. Those \NFTs are sometimes called \emph{real-time} in the literature.}.  
It is deterministic (written \DFT) if the underlying automaton is a \DFA.

We say that two transducers $T,T'$ are equivalent, denoted by $T\equiv T'$, whenever they define the same transduction, i.e. $R(T) = R(T')$. For all transducer classes $\mathcal{C}$, we say that
a transduction $R\subseteq \Sigma^*\times \Sigma^*$ is $\mathcal{C}$-definable if
there exists $T {\in} \mathcal{C}$ such that $R {=} R(T)$.
Given two classes $\mathcal{C},\mathcal{C'}$ of transducers, and a transducer $T\in\mathcal{C}$, we say
that $T$ is (effectively) $\mathcal{C}'$-definable if one can construct an equivalent transducer $T'\in\mathcal{C}'$. 

The \emph{$(\mathcal{C},\mathcal{C'})$-definability problem} takes as input a transducer $T\in\mathcal{C}$ and
asks to decide whether $T$ is $\mathcal{C'}$-definable. If so, one may want to construct
an equivalent transducer $T'\in \mathcal{C'}$. In this paper, we prove that $(\FUNTW, \NFT)$-definability is decidable.


It is known that whether an \NFT $T$ is functional can be decided
in PTime \cite{GurIba83}. The class of functional \NFTs is denoted by \FUNNFT. Functional \NFTs are strictly more expressive than
\DFTs. For instance, the function that maps any word $u\in\{a,b\}^+$ to
 $a^{|u|}$  if $u[|u|] = a$, and to  $b^{|u|}$ otherwise, is \FUNNFT-definable but not \DFT-definable.
This result does not hold for \TWNFTs: functional \TWNFTs and \TWDFTs define the same class
of transductions (Theorem 22 of \cite{EngHoo01}). 

\vspace{1mm}
\noindent\textbf{Examples}
  Let $\Sigma = \{a,b\}$ and $\#\not\in \Sigma$, and
  consider the transductions
  \begin{enumerate}
  \item $R_0 = \{ (u, a^{|u|})\ |\ u\in\Sigma^+, u[|u|]=a\}$
  \item $R_1 = \{ (u, b^{|u|})\ |\ u\in\Sigma^+, u[|u|]=b\}\cup R_0$
  \item $R_2 = \{ (\#u\#, \#\overline{u}\#)\ |\ u\in\Sigma^*\}$.
  \end{enumerate}

  $R_0$ is \DFT-definable: it suffices to replace each letter by $a$ and to accept only if the last letter is $a$. Therefore
  it can be defined by the \DFT $T_0 {=} (\{q_a,q_b\}, q_b, \{q_a\}, \{ (q_x,y,a,q_y)\ |\ x,y\in\Sigma\})$.

  $R_1$ is \FUNNFT-definable but not \DFT-definable: similarly as before we can define a \DFT $T'_0 = (\{ p_a,p_b\}, p_a, \{p_b\},\{ (p_x,y,b,p_y)\ |\ x,y\in\Sigma\}) $ that
  defines the transduction $\{ (u, b^{|u|})\ |\ u\in\Sigma^+, u[|u|]=b\}$, and construct an \NFT $T_1$ 
  as follows: its initial state is some fresh state $p_0$, and when reading $x\in\Sigma$ the first time, it non-deterministically
  goes to $T_0$ or $T'_0$ by taking the transition $(p_0,x,a,q_x)$ or $(p_0,x,b,p_x)$, and proceeds in either $T_0$ or
  $T'_0$. Even if $R_1$ is functional, it is not \DFT-definable, as the transformation depends on the property 
  of the last letter, which can be arbitrarily far away from the beginning of the string.

 $R_2$ is \TWDFT-definable: it suffices to
  go to the end of the word by producing $\epsilon$ each time a letter is read, to go back to the beginning
while copying each input letter, and return to the end without outputting anything, and to accept.
Hence it is defined by $T_2 = (\{ q_0, q_1,q_2,q_3,q_f\}, q_0, \{q_f\}, \delta_2)$ where states $q_1,q_2,q_3$ denote
passes, and $\delta_2$ is made
of the transitions $(q_0,\#,\epsilon,q_1,+1)$, $(q_1, x{\in}\Sigma, \epsilon, q_1, +1)$ (during the first pass, move to the right), 
$(q_1, \#, \epsilon, q_2, -1)$, $(q_2, x{\in}\Sigma,x, q_2, -1)$, $(q_2, \#, \#, q_3, +1)$, $(q_3, x{\in}\Sigma, \epsilon, q_3, +1)$,
$(q_3, \#, \#, q_f, +1)$.




\vspace{1mm}
\noindent\textbf{Crossing Sequences,  Loops and Finite-Crossing \TWNFTs}
The notion of crossing sequence is a useful 
notion in the theory of two-way automata \cite{Shepherdson59,hopcroft-ullman:1979a}, that
allows one to pump runs of two-way automata.
Given a \TWNFA $A$, a word $u\in\Sigma^*$ and a
run $\rho$ of $A$ on $u$, the \emph{crossing sequence} at position $i$, denoted by $\mbox{CS}(\rho,i)$ is given by the 
sequence of states $q$ such that $(q,i)$ occurs in $\rho$. The order of the sequence
is given by the order in which the pairs of the form $(q,i)$ occur in $\rho$. E.g.
if $\rho=(q_1,1)(q_2,2)(q_3,1)(q_4,2)(q_5,1)(q_6,2)(q_7,3)$ then
$\mbox{CS}(\rho, 1) = q_1q_3q_5$, $\mbox{CS}(\rho, 2) = q_2q_4q_6$ and $\mbox{CS}(\rho,3)=q_7$.
We write $\mbox{CS}(\rho)$ the sequence $\mbox{CS}(\rho,1),\dots,\mbox{CS}(\rho,|u|+1)$.

Crossing sequences allow one to define the loops of a run. Given a run
$\rho$ of the \TWNFA $A$ on some word $u$ of length $n$, a pair of
positions $(i,j)$ is a \emph{loop}~\footnote{Observe that we include
  the input letter in the notion of loop. We use this to avoid
  technical difficulties due to backward transitions (which do not
  read the local symbol, but its successor).} 
 in $\rho$ if $(i)$ $1\leq i\leq j\leq n$, $(ii)$
$\mbox{CS}(\rho,i) = \mbox{CS}(\rho,j)$ and $(iii)$ $u[i] = u[j]$. Let $u_1 =
u[1..(i-1)]$, $u_2=u[i..(j-1)]$ and $u_3 = u[j..n]$.  If $(i,j)$ is a
loop in $\rho$ and $u\in L(A)$, then $u_1(u_2)^ku_3\in L(A)$ for all
$k{\geq} 0$. We say that a loop $(i,j)$ is \emph{empty} if
$i=j$, in this case we have $u_2=\varepsilon$. %
The notions of crossing sequence and loop carry over to transducers
through their underlying input automata.

Given a \TWNFT $T$, $N\in\mathbb{N}$ and a run $\rho$ of $T$ on a word of length $n$, $\rho$ is said to 
be $N$-crossing if $|\mbox{CS}(\rho,i)|\leq N$ for all $i\in\{1,\dots, n\}$.
The transducer $T$ is \emph{finite-crossing} if there exists $N\in\mathbb{N}$ such that for all $(u,v)\in R(T)$, there is
an accepting $N$-crossing run $\rho$ on $u$ such that $\out(\rho) = v$. In that case,
$T$ is said to be $N$-crossing. It is easy to see that if $T$ is $N$-crossing, then for all $(u,v)\in R(T)$ 
there is an accepting run $\rho$ on $u$ such that $\out(\rho) = v$ and 
no states repeat in $\mbox{CS}(\rho,i)$ for all $i\in\{1,\dots,|u|\}$. Indeed, if some state $q$ repeats
in some $\mbox{CS}(\rho,i)$, then it is possible to pump the subrun between the two occurrences of $q$
on $\mbox{CS}(\rho,i)$. This subrun has an empty output, otherwise $T$ would not be functional.

\begin{proposition}\label{prop:ncrossing}
Any \FUNTW with $N$ states is $N$-crossing. 
\end{proposition}


%% file: generalcase.tex
\section{From Two-way to One-way Transducers}\label{sec:generalcase}

In this section, we prove the main result of this paper, i.e. the decidability of
$(\FUNTW, \NFT)$-definability. 
%
%

\subsection{Rabin and Scott's Construction for Automata}

The proof of Theorem \ref{thm:main} relies on the same ideas as Rabin and Scott's
construction for automata \cite{RabinScott59}. It is based on the following key observation:
Any accepting run is made of many zigzags, and those zigzags are organized by a nesting hierarchy: zigzag
patterns may be composed of simpler zigzag patterns. The simplest zigzags of the hierarchy
are those that do not nest any other zigzag: they are called \zmotions.
Rabin and Scott described a procedure that removes
those zigzags by iterating a construction that removes \zmotions. 
\input{figures/squeeze}

A \emph{one-step sequence} is an indexed sequence $s=a_1,\dots, a_n$ 
%
of positions such that $a_i\in \{1,2\dots, m\}$, 
$a_1=1$, $a_n=m$, and $|a_{i+1} - a_i| = 1$.
 The sequence $s$ is \emph{$N$-crossing}  if for all  $x\in \{1,2\dots, m\}$ we have  
 $|\{i \mid a_i= x\}|\leq N$. 
The reversals of $s$ are the indexes $1<r_1 < r_2 < \dots <r_l <n$ such that 
 $a_{r_i +1}= a_{r_i-1} $. In the sequel we let $r_0 = 1$ and $r_{l+1}=n$.

 A \emph{\zmotion} $z$ in $s$ is a subsequence $a_{e}, a_{e +
   1},\dots a_{f}$ such that there is $0< i < l$ with $r_{i-1}\leq e
 <r_{i} < r_{i+1}< f \leq r_{i+2}$, and $a_{e}=a_{r_{i+1}}$ and
 $a_{f}=a_{r_{i}}$.  We may denote $z$ by the pair of reversals
 $(r_i,r_{i+1})$.  E.g. the sequences $z_1 = 1,2,3,2,1,2,3$ and $z_2 =
 4,3,2,3,4,3,2$ are \zmotions.  The \emph{shape} of a run $\rho$ is
 defined as the second projection of $\rho$, written $\shape(\rho)$. A
 run $\rho$ is a \emph{\zmotion run} if $\shape(\rho)$ is a
 \zmotion. When there is no ambiguity,
 \zmotion runs are just called \zmotions.



If $T$ is a \TWNFA, it is possible to construct a new automaton denoted by $\squeeze{T}$ such that, 
for all accepting runs $\rho$ of $T$ on some input word $u$,  there exists a ``simpler'' accepting run of
$\squeeze{T}$ on $u$,  obtained from $\rho$ by replacing some \zmotions by one-way runs
that simulate three passes in parallel. It is illustrated by Fig.~\ref{fig:squeeze}. 
For instance at the first step, 
there are two \zmotions from $q_1$ to $q_2$ and from $q_3$ to $q_4$ respectively. Applying $\squeeze{T}$ consists in non-deterministically guessing those \zmotions and simulating them by one-way runs.
This is done by the \NFA $R_T(q_1,q_2)$ and 
$R_T(q_3,q_4)$ respectively. Depending on whether the \zmotions enter from the left or the right, 
\zmotions are replaced by runs of \NFAs $R_T(.,.)$ (that read the input backwardly) or $L_T(.,.)$ , as illustrated by the
second iteration of $\squeezeOp$ on Fig.~\ref{fig:squeeze}.

An $N$-crossing run $\rho$ can be simplified into a one-way run after a constant number of
applications of $\squeezeOp$. This result is unpublished so we prove it in this paper.
In particular, we show that if $\rho$ is $N$-crossing, 
then its zigzag nesting depth decreases after $N$ steps. Moreover, if $\rho$ is $N$-crossing, then
its zigzag nesting depth is also bounded by $N$. Therefore after $N^2$ applications of 
$\squeezeOp$, $\rho$ is transformed into a simple one-way run. It is sufficient to prove those
results at the level of integer sequences. In particular, one can define $\squeeze{s}$ the set of
sequences obtained from a one-step sequence $s$ by replacing \emph{some} \zmotions of $s$
by strictly increasing or decreasing subsequences. The following is formalized and shown in Appendix:
\begin{lemma}\label{lem:removalsquare}
Let $s$ be an $N$-crossing one-step sequence over $\{1, \dots , m\}$.
Then $1,2, \dots, m$ is in $\squeezepow{N^2}{s}$.
\end{lemma}

At the automata level, it is known that for all words $u$ accepted by a \TWNFA $T$ with $N$ states, there
exists an $N$-crossing accepting run on $u$. Therefore it suffices to apply
$\squeezeOp$ $N^2$ times 
to $T$. One gets an equivalent \TWNFA $T^*$ from which the backward transitions can be removed 
while preserving equivalence with $T^*$, and so $T$.

\subsection{Extension to transducers: overview}

The construction used to show decidability of \NFT-definability of  \FUNTW follows the same 
ideas as Rabin and Scott's construction. The main difference relies in the transformation
of the local transducers defined by \zmotion runs (that we call \ZNFTs) into
\NFTs. Our procedure is built over a \ZNFT-to-\NFT procedure. It is seen as a black-box in this section,
but is the subject of the next section.

Compared to two-way automata, one faces an extra difficulty caused by the fact that \TWNFTs (and \ZNFTs) are not always \NFT-definable. Therefore one defines a necessary condition that has to be tested each time we want to apply $\squeezeOp$. Let us consider again Fig.~\ref{fig:squeeze} when $T$ is a \TWNFT. 
One defines from $T$ the transductions induced by local \zmotion runs from a starting state $q_1$ to an ending state $q_2$, 
and show that those local transductions must be \NFT-definable.

Once this necessary condition is satisfied, the construction $\squeezeOp$ can be applied and works as for Rabin and Scott's construction:
the new transducer $\squeeze{T}$ simulates $T$ and non-deterministically 
may guess that the next zigzag of $T$ is a \zmotion run from some
state $q_1$ to some state $q_2$, and thus can be simulated by a run of some
\NFT $R_T(q_1,q_2)$ or $L_T(q_1,q_2)$, depending on whether it
 enters from the left or the right. Then $\squeeze{T}$
switches to $R_T(q_1,q_2)$ (if it entered from the right)
and once $R_T(q_1,q_2)$ reaches an accepting state, it may come back to its normal mode. 

\subsection{\zmotion transducers}

\zmotion transducers are defined like \TWNFTs except that they must define \textbf{functions} and
to be accepting, a run on a word of length $n$ must be of the form $\rho.(q_f,n+1)$ where $\rho$
is a \zmotion run and $q_f$ is an accepting state. Note that it implies that 
$\shape(\rho)$ is always of the form $1,\dots,n,n{-}1,\dots,1,\dots,n$.
The class of \zmotion transducers is denoted by \ZNFTs. Note that \zmotion transducers are incomparable with
\FUNTWs. Indeed, \zmotion transducers can define the transduction
$u\in\Sigma^*\mapsto \overline{u}$, which is not \FUNTW-definable as there are no
end markers. %
%
%
%
%

Let $T\in\ZNFT$ and $\rho = (p_1,1)\dots (p_n,n)$
$(q_{n-1},n{-}1)\dots (q_1,1)(r_2,2)\dots (r_{n+1},n+1)$ be a run of
$T$ on a word of length $n$. We let $q_n = p_n$ and $r_1 = q_1$ and
define the following shortcuts: for $1\leq i\leq j\leq n$,
$\out_1[i,j] = \out((p_i,i)\dots (p_j,j))$, and $\out_2[i,j] =
\out((q_j,j)\dots (q_i,i))$ and $\out_3[i,j] = \out((r_i,i)\dots
(r_j,j))$, and $\out_3[i,n+1] = \out((r_i,i)\dots (r_{n+1},n+1))$.

\input{figures/property_P}


We characterize the \NFT-definability of a \ZNFT by a property that we
prove to be decidable. Intuitively, this property requires that the
outputs produced by loops can be produced by a single forward pass:

\begin{definition}[$\ppt$-property]
    Let $T$ be  a \ZNFT. We say that $T$ satisfies the property $\ppt$, denoted
    by $T\models \ppt$, if for all words $u\in \dom(T)$, for all 
    accepting runs $\rho$ 
    on $u$, and for all pairs of loops 
  $(i_1,j_1)$ and $(i_2,j_2)$ of $\rho$ such that $j_1\leq i_2$, 
    there exist $\beta_1,\beta_2,\beta_3,\beta_4,\beta_5\in \Sigma^*$, 
    $f,g:\mathbb{N}^2\rightarrow \Sigma^*$ and constants
    $c_1,c'_1,c_2,c'_2 \geq 0$ such that $c_1,c_2\neq 0$ and for all $k_1,k_2\geq 0$, 
    $$\begin{array}{c}
    f(k_1,k_2)x_0v_1^{\eta_1}x_1w_1^{\eta_2}x_2w_2^{\eta_2}x_3v_2^{\eta_1}x_4v_3^{\eta_1}x_5w_3^{\eta_2}x_6g(k_1,k_2)\\
    = \beta_1\beta_2^{k_1}\beta_3\beta_4^{k_2}\beta_5
    \end{array}
    $$
    where $\eta_i=k_ic_i+c'_i$, $i\in\{1,2\}$, and, $x_i$'s, $v_i$'s
    and $w_i's$ are words defined as depicted in Fig.~\ref{fig:ppt}.
\end{definition}

The following key lemma is proved in Section~\ref{sec:zmotions}.

\begin{lemma}\label{lem:characterization}
    Let $T\in \ZNFT$. $T\models \ppt$ iff $T$ is \NFT-definable. Moreover, 
    $\ppt$ is decidable and if $T\models \ppt$, one can (effectively) construct
    an equivalent \NFT.
\end{lemma}


\begin{definition}[\zmotion transductions induced by a \FUNTW]\label{def:inducedZmotions}
Let $T = (Q,q_0,F,\Delta)$ be a \FUNTW and $q_1,q_2\in Q$.  The transduction $\lc_T(q_1,q_2)$ (resp. $\rc_T(q_1,q_2)$) is defined
as the set of pairs $(u_2,v_2)$ such that there exist $u\in\Sigma^*$, two positions $i_1<i_2$ (resp. $i_2<i_1$), an accepting run $\rho$ of $T$ on $u$ which
can be decomposed as $\rho = \rho_1(q_1,i_1)\rho_2(q_2,i_2)\rho_3$ such that $u_2 = u[i_1\dots i_2]$ and

\begin{itemize}
\item $(q_1,i_1)\rho_2(q_2,i_2)$ is a \zmotion run
\item $\out((q_1,i_1)\rho_2(q_2,i_2)) = v_2$
\end{itemize}
\end{definition}


\zmotions can be of two forms: either they start from the left and end to
the right, or start from the right and end to the left. In order to avoid considering
these two cases each time, we introduce the notation $\overline{T}$ 
that denotes the mirror of $T$: it is $T$ where the moves $+1$ are replaced by $-1$ and
the moves $-1$ by $+1$. Moreover, the way $\overline{T}$ reads the input tape is slightly modified: it starts
in position $n$ and a run is accepting if
it reaches position $0$ in some accepting state. All the notions defined for
\TWNFTs carry over to their mirrors. In particular, $(u,v)\in R(T)$ iff $(\overline{u}, v)\in R(\overline{T})$.
The \zmotion transductions $\rc_T(q_1,q_2)$ and $\lc_T(q_1,q_2)$ are symmetric in the following
sense: $\rc_T(q_1,q_2) = \lc_{\overline{T}}(q_1,q_2)$ and $\lc_T(q_1,q_2) = \rc_{\overline{T}}(q_1,q_2)$. 

\begin{proposition}\label{prop:localzmotions}
    The transductions $\rc_T(q_1,q_2)$ and $\lc_T(q_1,q_2)$ are \ZNFT-definable.
\end{proposition}

\begin{IEEEproof}
We only consider the case $\lc_T(q_1,q_2)$, the other case being solved by using the equality
$\rc_T(q_1,q_2) = \lc_{\overline{T}}(q_1,q_2)$. We first construct from $T$ a \ZNFT $Z'_T(q_1,q_2)$
which is like $T$ but its initial state is $q_1$, and it can move to an accepting
state whenever it is in $q_2$. However $Z'_T(q_1,q_2)$ may define input/output pairs
$(u_2,v_2)$  that cannot be embedded into some pair $(u,v)\in R(T)$ as required by the definition of 
$\lc_T(q_1,q_2)$. Based on Shepherdson's construction, we modify $Z'_T(q_1,q_2)$ in order
to take this constraint into account. The full proof is in Appendix.
\end{IEEEproof}

In the next subsection, we show that $\rc_T(q_1,q_2)$ and $\lc_T(q_1,q_2)$ must necessarily
be \NFT-definable for $T$ to be \NFT-definable. For that purpose, 
it is crucial in Definition~\ref{def:inducedZmotions} to
make sure that the \zmotion $(q_1,i_1)\rho_2(q_2,i_2)$ can be embedded
into a global accepting run of $T$.  Without that restriction, it
might be the case that $\lc_T(q_1,q_2)$ or $\rc_T(q_1,q_2)$ is not
\NFT-definable although the \TWNFT $T$ is. Indeed, the domain of
$\lc_T(q_1,q_2)$ or $\rc_T(q_1,q_2)$ would be too permissive and
accept words that would be otherwise rejected by other passes of
global runs of $T$. This is another difficulty when lifting Rabin and
Scott's proof to transducers, as for automata, the context in which a
\zmotion occurs is not important.

\subsection{Decision procedure and proof of Theorem \ref{thm:main}}

We show that the construction $\squeeze{T}$ can be applied if the following
necessary condition is satisfied.

\begin{lemma}\label{lem:gen-cns}
  If $T$ is \NFT-definable, then so are the transductions
  $\rc_T(q_1,q_2)$ and $\lc_T(q_1,q_2)$ for all states $q_1,q_2$.
  Moreover, it is decidable whether the transductions $\rc_T(q_1,q_2)$
  and $\lc_T(q_1,q_2)$ are \NFT-definable.
\end{lemma}

\begin{IEEEproof}[Sketch of proof]
  We have seen in Lemma~\ref{lem:characterization} that
  \NFT-definability of an \ZNFT is characterized by Property
  $\ppt$.
  Let $Z\in\ZNFT$ that defines $\lc_T(q_1,q_2)$ for some $q_1,q_2$, we
   thus sketch the proof that $Z\models \ppt$.

  Consider two loops $(i_1,j_1)$, $(i_2,j_2)$ of a run $\rho$ of $Z$
  on some word $u$, as in the premises of Property $\ppt$. They
  induce a decomposition of $u$ as $u=u_1u_2u_3u_4u_5$ with
  $u_2=u[i_1\dots j_1-1]$ and $u_4=u[i_2\dots j_2-1]$. By definition of the
  transduction $\lc_T(q_1,q_2)$, any word in $\dom(Z)$ can be extended
  into a word in $\dom(T)$. By hypothesis, $T$ is \NFT-definable, thus
  there exists an equivalent \NFT $T'$. As $T'$ has finitely many
  states, it is possible, by iterating the loops $(i_1,j_1)$ and
  $(i_2,j_2)$, to identify an input word of the form 
  $$u' = \alpha u_1 u_2^{c_1} u_2^{c_2} u_2^{c_3} u_3 u_4^{c'_1} u_4^{c'_2} u_4^{c'_3} u_5 \alpha'$$
  and a run $\rho'$ of $T'$ on this word which has two loops on the
  input subwords $u_2^{c_2}$ and $u_4^{c'_2}$. It is then easy to
  conclude.
\end{IEEEproof}

\vspace{2mm}
\noindent \textbf{Construction of $\squeeze{T}$}
Assuming that the necessary condition is satisfied, we now explain how to construct 
the \FUNTW $\squeeze{T}$. By hypothesis, the transductions $\lc_T(q_1,q_2)$ and
$\rc_T(q_1,q_2)$ are \NFT-definable for all $q_1,q_2$ by \NFT $L_T(q_1,q_2)$ and
$R_T(q_1,q_2)$ respectively (they exist by Proposition \ref{prop:localzmotions} and 
Lemma~\ref{lem:characterization}). As already
said before, the main idea to define $\squeeze{T}$ is to non-deterministically (but repeatedly) apply
$L_T(q_1,q_2)$, $R_T(q_1,q_2)$, or $T$, for some $q_1,q_2\in Q$. However when applying
$R_T(q_1,q_2)$, the head of $\squeeze{T}$ should move from the right to the left, so that
we have to mirror the transitions of $R_T(q_1,q_2)$.

The transducer $\squeeze{T}$ has two modes, \Zmode or \Tmode. 
In \Tmode, it works as $T$ until it non-deterministically decides that the
next zigzag is a \zmotion from some state $q_1$ to some state $q_2$. Then it goes in
\Zmode and runs $L_T(q_1,q_2)$ or $\overline{R_T(q_1,q_2)}$, in which transitions to
an accepting state have been replaced by transitions from $q_2$ in $T$, so that
$\squeeze{T}$ returns in \Tmode. From those transitions we also add transitions from the initial states of
$L_T(q_2,q_3)$  and $\overline{R_T(q_2,q_3)}$ for all $q_3\in Q$, in case $\squeeze{T}$ guesses
that the next \zmotion starts 
immediately at the end of the previous \zmotion. 
We detail the construction of $\squeeze{T}$ in Appendix.

\begin{proposition}\label{prop:correctness}
Let $T\in\FUNTW$ such that $T$ is \NFT-definable.
Then $\squeeze{T}$ is defined and equivalent to $T$.
\end{proposition}

Let $T\in \FUNTW$. If $T$ is \NFT-definable, then the operator 
$\squeezeOp$ can be iterated on $T$ while preserving
equivalence with $T$, by the latter proposition. By Proposition
\ref{prop:ncrossing} $T$ is $N$-crossing, and therefore,
based on Lemma \ref{lem:removalsquare}, it suffices to iterate
$\squeezeOp$ $N^2$ times to remove all zigzags from accepting runs of $T$,
as stated by the following lemma:

\begin{lemma}
    Let $T$ be a \FUNTW with $N$ states. If $T$ is \FUNNFT-definable, then 
    $\squeezepow{N^2}{T}$ is defined and equivalent to $T$, and moreover, for all $(u,v)\in R(T)$, 
    there exists an accepting run $\rho$ of $\squeezepow{N^2}{T}$ on $u$ such that 
    $\out(\rho) = v$ and $\rho$ is made of forward transitions only.
\end{lemma}


\noindent\textbf{Proof of Theorem \ref{thm:main}} In order to decide whether a $\FUNTW$ $T$ is \NFT-definable, it suffices
to test whether $\squeezeOp$ can be applied $N^2$ times. More precisely, 
it suffices to set $T_0$ to $T$, $i$ to $0$, and, while $T_i$ satisfies the necessary condition (which is
decidable by Lemma \ref{lem:gen-cns}) and $i\leq N^2$, to increase $i$ and set
$T_i$ to $\squeeze{T_{i-1}}$. If the procedure exits the loops before reaching $N^2$, 
then $T$ is not \NFT-definable, otherwise it is \NFT-definable by the \NFT 
obtained by removing from $T_{N^2}$ all its backward transitions.

%% file: figures/squeeze.tex
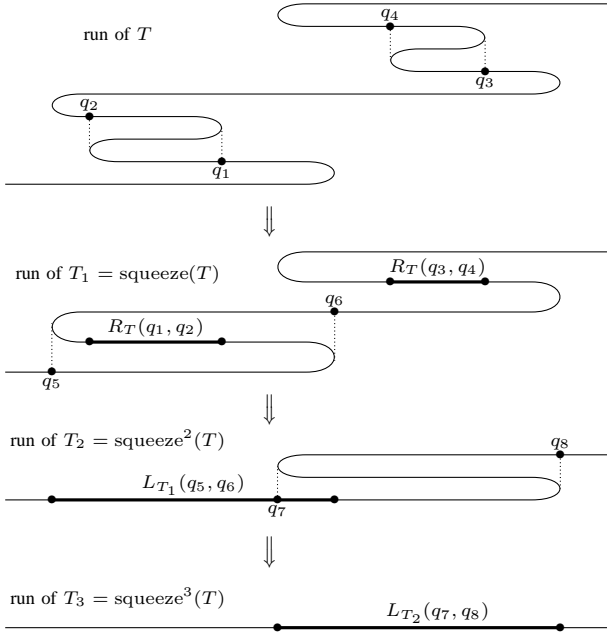
\begin{figure}[t]
   \centering
   \scriptsize
   \begin{tikzpicture}[]


     \draw 
       (0,0)     -- (4,0)     .. controls (4.5,0)   and (4.5,.3) ..
       (4,.3)    -- (1.5,.3)  .. controls (1,.3)    and (1,.6) ..
       (1.5,.6)  -- (2.5,.6)  .. controls (3,.6)    and (3,.9) ..
       (2.5,.9)  -- (1,.9)    .. controls (.5,.9)   and (.5,1.2) ..
       (1,1.2)   -- (7,1.2)   .. controls (7.5,1.2) and (7.5,1.5) ..
       (7,1.5)   -- (5.5,1.5) .. controls (5,1.5)   and (5,1.8) ..
       (5.5,1.8) -- (6,1.8)   .. controls (6.5,1.8) and (6.5,2.1) ..
       (6,2.1)   -- (4,2.1)   .. controls (3.5,2.1) and (3.5,2.4) ..
       (4,2.4)   -- (8,2.4)
     ;

     \draw (2.88,.15) node {$q_1$};
     \draw[densely dotted] (2.88,.3) node {$\bullet$} -- (2.88,.75);

     \draw (1.12,1.05) node {$q_2$};
     \draw[densely dotted] (1.12,.9) node {$\bullet$} -- (1.12,.45);

     \draw (6.38,1.35) node {$q_3$};
     \draw[densely dotted] (6.38,1.5) node {$\bullet$} -- (6.38,1.95);

     \draw (5.12,2.25) node {$q_4$};
     \draw[densely dotted] (5.12,2.1) node {$\bullet$} -- (5.12,1.65);

     \draw (1.5,2) node {run of $T$};

     \draw (3.5,-.5) node {\rotatebox{270}{$\implies$}};


     \begin{scope}[yshift=-2.5cm]
     \draw 
       (0,0)   -- (4,0)   .. controls (4.5,0)  and (4.5,.4) ..
       (4,.4)  -- (1,.4)  .. controls (.5,.4)  and (.5,.8) ..
       (1,.8)  -- (7,.8)  .. controls (7.5,.8) and (7.5,1.2) ..
       (7,1.2)  -- (4,1.2)  .. controls (3.5,1.2) and (3.5,1.6) ..
       (4,1.6) -- (8,1.6)
     ;

     \draw (2,.6) node {$R_T(q_1,q_2)$};
     \draw[very thick] (2.88,.4) node {$\bullet$} -- (1.12,.4) node {$\bullet$};

     \draw (5.75,1.4) node {$R_T(q_3,q_4)$};
     \draw[very thick] (6.38,1.2) node {$\bullet$} -- (5.12,1.2) node {$\bullet$};

     \draw (.62,-.15) node {$q_5$};
     \draw[densely dotted] (.62,0) node {$\bullet$} -- (.62,.65);

     \draw (4.38,.95) node {$q_6$};
     \draw[densely dotted] (4.38,.8) node {$\bullet$} -- (4.38,.15);

     \draw (1.5,1.3) node {run of $T_1 = \squeeze T$};

     \draw (3.5,-.5) node {\rotatebox{270}{$\implies$}};
     \end{scope}


     \begin{scope}[yshift=-4.2cm]
     \draw 
       (0,0)    -- (7,0)  .. controls (7.5,0)  and (7.5,.3) ..
       (7,.3)   -- (4,.3) .. controls (3.5,.3) and (3.5,.6) ..
       (4,.6) -- (8,.6)
     ;

     \draw (2.5,.2) node {$L_{T_1}(q_5,q_6)$};
     \draw[very thick] (.62,0) node {$\bullet$} -- (4.38,0) node {$\bullet$};

     \draw (3.62,-.15) node {$q_7$};
     \draw[densely dotted] (3.62,0) node {$\bullet$} -- (3.62,.45);

     \draw (7.38,.75) node {$q_8$};
     \draw[densely dotted] (7.38,.6) node {$\bullet$} -- (7.38,.15);

     \draw (1.5,.8) node {run of $T_2 = \squeezepow 2 T$};

     \draw (3.5,-.7) node {\rotatebox{270}{$\implies$}};
     \end{scope}


     \begin{scope}[yshift=-5.9cm]
     \draw 
       (0,0) -- (8,0)
     ;

     \draw (5.75,.2) node {$L_{T_2}(q_7,q_8)$};
     \draw[very thick] (3.62,0) node {$\bullet$} -- (7.38,0) node {$\bullet$};

     \draw (1.5,.4) node {run of $T_3 = \squeezepow 3 T$};
     \end{scope}

   \end{tikzpicture}
\vspace{-2mm}
   \caption{Zigzags removal by applications of $\squeezeOp$.
     \label{fig:squeeze}}
\vspace{-7mm}
\end{figure}

%% file: figures/property_P.tex
\begin{figure}[t]
   \centering
   \scriptsize
   \begin{tikzpicture}[]

     \draw [->,>=latex] 
       (0,0)    -- (7,0)  .. controls (7.7,0)  and (7.7,.8) ..
       (7,.8)   -- (.5,.8) .. controls (-.2,.8) and (-.2,1.6) ..
       (.5,1.6) -- (7.5,1.6) ;

     \draw [->]  [bend right]      (7.7,.1)   -- (7.7,.7);

     \draw[densely dotted] (1.4,-.2) -- (1.4,1.8);
     \draw[densely dotted] (2.8,-.2) -- (2.8,1.8);
     \draw[densely dotted] (4.2,-.2) -- (4.2,1.8);
     \draw[densely dotted] (5.6,-.2) -- (5.6,1.8);

      \draw (0,0) node {$\bullet$};

     \draw (1.4,-.4) node {$i_1$};
     \draw (2.8,-.4) node {$j_1$};
     \draw (4.2,-.4) node {$i_2$};
     \draw (5.6,-.4) node {$j_2$};

     \draw (.7,.15) node {$x_0$};
     \draw (2.1,.15) node {$v_1$};
     \draw (3.5,.15) node {$x_1$};
     \draw (4.9,.15) node {$w_1$};
     \draw (6.8,.4) node {$x_2$};

     \draw (.7,1.15) node {$x_4$};
     \draw (2.1,.95) node {$v_2$};
     \draw (3.5,.95) node {$x_3$};
     \draw (4.9,.95) node {$w_2$};

     \draw (2.1,1.75) node {$v_3$};
     \draw (3.5,1.75) node {$x_5$};
     \draw (4.9,1.75) node {$w_3$};
     \draw (6.8,1.75) node {$x_6$};

   \end{tikzpicture}
   \vspace{-10pt}
   \caption{Output decomposition in property $\ppt$.
     \label{fig:ppt}}
   \vspace{-10pt}
\end{figure}

%% file: zmotions.tex
\section{From Elementary Zigzags to Lines}\label{sec:zmotions}

This section is devoted to the proof of Lemma~\ref{lem:characterization} that
characterizes \NFT-definable \ZNFT by the property $\ppt$ and states its
decidability. Moreover, we give a \ZNFT-to-\NFT construction when $\ppt$
is satisfied.

We first prove that Property $\ppt$ is a necessary condition for
\NFT-definability. To prove the converse, we proceed in two
steps.
First, we define a procedure that tests whether a given \ZNFT $T$ is
equivalent to a \ZNFT that does not output anything on its backward
pass (called \eZNFT), and then define another procedure that tests
whether the latter \ZNFT is equivalent to an \NFT. We show that it is
always true whenever $T\models \ppt$. This approach is depicted in
Fig.~\ref{fig:ZNFT2NFT}. The two steps are similar, therefore we
mainly focus on the first step.

\input{figures/znft-to-nft}

\subsection{Property $\ppt$ is a necessary condition}

We show that Property $\ppt$ only depends on transductions.
\begin{lemma}\label{lem:ppt-equiv}
Let $T,T'{\in}\ZNFT$. If $T{\models} \ppt$ and $T{\equiv} T'$ then $T'{\models} \ppt$.
\end{lemma}

\begin{IEEEproof}
  Consider two loops $(i_1,j_1)$, $(i_2,j_2)$ as in Property $\ppt$ in
  a run of $T'$ on some word $u$.  They induce a decomposition of $u$
  as $u=u_1u_2u_3u_4u_5$ where $u_2=u[i_1\dots(j_1-1)]$ and
  $u_4=u[i_2\dots(j_2-1)]$, with $u_1u_2^{k_1}u_3u_4^{k_2}u_5
  \in \dom(T')$ for all $k_1,k_2\geq 0$.

  As $T$ is equivalent to $T'$ and has finitely many states, there
  exist iterations of the loops on $u_2$ and $u_4$ which constitute
  loops in $T$ on powers of $u_2$ and $u_4$. Formally, there exist
  integers $d_1,e_1,h_1,d_2,e_2,h_2$ with $e_1,e_2>0$ such that $T$
  has a run $\rho$ on the input word
  $u_1u_2^{d_1}u_2^{e_1}u_2^{h_1}u_3u_4^{d_2}u_4^{e_2}u_4^{h_2}u_5$
  which contains a loop on the input subwords $u_2^{e_1}$ and
  $u_4^{e_2}$.

  We conclude easily by using the fact that $T\models\ppt$.
%
\end{IEEEproof}

As a consequence, we obtain that Property $\ppt$ is a necessary
condition for \NFT-definability.
\begin{lemma}\label{lem:z-cns}
    Let $T\in \ZNFT$. If $T$ is \NFT-definable, then $T\models \ppt$.
\end{lemma}
\begin{IEEEproof}
  Let $T'$ be an \NFT equivalent to $T$. It is easy to turn $T'$ into
  a \ZNFT $T''$ that performs two additional backward and forward
  passes which output $\varepsilon$. Consider two loops $(i_1,j_1)$
  and $(j_1,j_2)$ in a run of $T''$, and let us write the output of
  this run as depicted on Fig.~\ref{fig:ppt}. These loops are also
  loops of $T'$, and thus we can define $\beta_1$ (resp. $\beta_2$,
  $\beta_3$, $\beta_4$ and $\beta_5$) as $x_0$ (resp. $v_1$, $x_1$,
  $w_1$ and $x_2$), and $f,g$ as the constant mappings equal to
  $\epsilon$. Hence $T''\models \ppt$, and we conclude by
  Lemma~\ref{lem:ppt-equiv}.
\end{IEEEproof}

\subsection{From \ZNFT to \eZNFT}
The goal 
is to devise a procedure that tests whether the first and second
passes (forward and backward) of the run can be done with a single
forward pass, and constructs an \NFT that realizes this single forward
pass. Then, in order to obtain an \eZNFT, it suffices to replace the
first pass of $T$ by the latter \NFT and add a backward pass that just
comes back to the beginning of the word and outputs $\epsilon$ all the
time. The procedure constructs an \eZNFT, and tests whether it is
equivalent to $T$. It is based on the following key property that
characterizes the form of the output words of the two first passes of any
\ZNFT satisfying $\ppt$.
 Intuitively, when these words are long
enough, they can be decomposed as words whose primitive roots are
conjugate.



\input{figures/property_P1}

\begin{definition}[$\pptone$-property]
  Let $T \in \ZNFT$ with $m$ states, and let $(u,v)\in R(T)$ where $u$
  has length $n$. Let $K = 2.o.m^3.|\Sigma|$ where $o = max\{|v|\ |\
  (p,a,v,q,m)\in \Delta\}$.  The pair $(u,v)$ satisfies the property
  $\pptone$, denoted by $(u,v)\models \pptone$, if for all accepting
  runs $\rho$ on $u$, 
  there exist a position $1\leq \ell\leq n$ and
  $w,w',t_1,t_2,t_3\in\Sigma^*$ such that $v \in wt_1t_2^*t_3w'$ and:
$$
\begin{array}{@{}l@{}ll}
\out_1[1,\ell] = w  & \out_2[1,\ell] = t_3 & \out_1[\ell, n]\out_2[\ell,n] {\in} t_1t_2^*\\
\out_3[1,n+1]  = w' & & |t_i|\leq 2K, \forall i\in \{1,2,3\}
\end{array}
$$
        This decomposition is depicted in Fig.~\ref{fig:pptone}.
    $T$ satisfies property $\pptone$, denoted $T\models \pptone$, if all $(u,v)\in R(T)$ satisfy it.
\end{definition}

\begin{proposition}\label{prop:outputsnecessary}
    Let $T\in \ZNFT$. If $T\models \ppt$, then $T\models \pptone$.
\end{proposition}


\begin{IEEEproof}
  \noindent $\bullet$ If $|\out_2[1,n-1]| \leq K$, then clearly, it
  suffices to take $\ell = n$, $t_1 = \out_2[n-1,n]$,
  $t_2=\varepsilon$, $t_3 = \out_2[0,n-1]$, $w = \out_1[1,n]$ and $w'
  = \out_3[1,n+1]$.


\noindent $\bullet$ Otherwise, $|\out_2[1,n-1]|>K$. Therefore $u$ is
of length $2.m^3.|\Sigma|$ at least and there exists a
(non-empty) loop $(i,j)$ in $\rho$.  We can always choose this loop
such that $|\out_2[1,i]| \leq K$ and $1\leq |\out_2[i,j]|\leq K$ (see
Lemma \ref{lem:simpleloopsruns} in Appendix).

\input{figures/loop}
The loop partitions the input and output words into factors that are
depicted in Fig. \ref{fig:zmotionrun} (only the two first passes are
depicted). Formally, let $u = u_1u_2u_3$ such that $u_2 =
u[i\dots(j{-}1)]$.  Let $x_0 = \out_1[1,i]$, $v_1 = \out_1[i,j]$, $x_1 =
\out_1[j,n]\out_2[j,n]$, $v_2 = \out_2[i,j]$, $x_2 = \out_1[1,i]$,
$x_3 = \out_3[1,i]$, $v_3 = \out_3[i,j]$ and $x_4 = \out_3[j,n+1]$. In
particular, we have $|x_2|\leq K$, $1\leq |v_2| \leq K$ and
$x_0v_1x_1v_2x_2x_3v_4x_4\in T(u)$. Since $(i,j)$ is a loop we also
get $x_0v_1^kx_1v_2^kx_2x_3v_3^kx_4\in T(u_1u_2^ku_3)$ for all $k\geq
0$. We then distinguish two cases:\\
\indent 1) If $v_1 \neq \epsilon$. We can apply Property $\ppt$ by taking
the second loop empty. We get that
  for all $k\geq 0$
  $$
  f(k)x_0v_1^{kc+c'}x_1v_2^{kc+c'}x_2x_3v_3^{kc+c'}x_4g(k) =
  \beta_1\beta_2^k\beta_3
  $$
  where $f,g:\mathbb{N}\rightarrow \Sigma^*$, $c\in\mathbb{N}_{>0}$,
  $c'\in\mathbb{N}$, and $\beta_1,\beta_2,\beta_3\in\Sigma^*$.
  Since the above equality holds for all $k\geq 0$, we can apply Lemma
  \ref{lem:fundamental} and we get $\primroot(v_1)\sim
  \primroot(\beta_2)$ and $\primroot(\beta_2) \sim \primroot(v_2)$,
  and therefore $\primroot(v_1) \sim \primroot(v_2)$. So there exist
  $x,y \in\Sigma^*$ such that
  $v_1\in (xy)^*$ and $v_2\in (yx)^*$. One can show (see Lemma \ref{lem:outputform} in Appendix)
  that $v_1x_1v_2\in x(yx)^*$. Then it suffices to take $\ell =
    i$, $w = x_0$, $t_1 = x$, $t_2 = yx$ and $t_3 = x_2$.
    
  \indent 2) The second case ($v_1=\epsilon$) is more complicated as it
    requires to use the full Property $\ppt$, using two non-empty
    loops.
    First, we distinguish two
    cases whether $|\out_1[j,n]|\leq K$ or not. For the latter case,
    we identify a second loop and then apply Property
    $\ppt$. Details can be found in the Appendix~\ref{app:zmotions}.
\end{IEEEproof}

\vspace{3mm}
\noindent\textbf{Construction of an \eZNFT from a \ZNFT} We construct an \eZNFT $T'$ from a \ZNFT $T$
such that $R(T') = \{ (u,v)\in R(T)\ |\ (u,v)\models
\pptone\}$. Intuitively, the main idea is to perform the two first
passes in a single forward pass, followed by a non-producing backward
pass, and the final third pass is exactly as $T$ does.  Therefore,
$T'$ guesses the words $t_1,t_2$ and $t_3$ and makes sure that the
output $v$ is indeed of the form characterized by $\pptone$. This can
be done in a one-way fashion while simulating the forward and backward
passes in parallel and by guessing non-deterministically the position
$\ell$. In addition, the output mechanism of $T'$ exploits the special
form of $v$: the idea is to output powers of $t_2$ while simulating
the two first passes.

First, let us describe how $T'$ simulates the forward and
backward passes in parallel during the first forward pass. 
It guesses both the state of the backward pass, and the
current symbol (this is needed as the symbol read by the backward
transition is the next symbol).  The first state ($q^*$) guessed for
the backward pass needs to be stored, as the last (forward) pass
should start from $q^*$. The transducer can go from state
$(p,q,\sigma)$ to state $(p',q',\sigma')$ if the current symbol is
$\sigma$ and there is a (forward) transition
$(p,\sigma,x,p',+1)$ and a (backward) transition
$(q',\sigma',y,q,-1)$.
Therefore if $Q$ is the
set of states of $T$, $T'$ uses, on the first pass, elements of
$Q\times Q \times \Sigma$ in its states. 
The transducer $T'$ can non-deterministically decide to perform the
backward and non-producing backward pass whenever it is in some state
$(q,q,\sigma)$ and the current symbol is $\sigma$. This indeed happens precisely when the forward and backward passes are in the same state $q$. 
If the current symbol is not the last of the input word, then the
whole run of $T'$ is not a \zmotion and therefore it is not
accepting.


Second, we describe how the \eZNFT $T'$, with the guess of
$t_1,t_2,t_3$, verifies during its first forward pass that the output
has the expected form, and how it produces this output.
%
%
%
During the first pass, $T'$ can be in two modes: In mode $1$ (before
the guess $\ell$), $T'$ verifies that the output on the simulated
backward pass is $t_3$ and proceeds as $T$ in the first forward pass
(it outputs what $T$ outputs on the forward pass).  Mode $2$ starts
when the guess $\ell$ has been made. In this mode, $T'$ first outputs
$t_1$ and then verifies that the output of the forward/backward run
from and to position $\ell$ is of the form $t_1t_2^*$. It can be done
by using pointers on $t_1$ and $t_2$.  There are two cases (guessed by
$T'$): either $t_1$ ends during the forward pass or during the
backward pass (using notations of Fig.\ref{fig:pptone}, either $t_1$
is a prefix of $x$, or $x$ is a prefix of $t_1$).

In the first case, $T'$ needs a pointer on $t_1$ to make sure that the
output of $T$ in the forward pass starts with $t_1$. It also needs a
pointer on $t_2$, initially at the end of $t_2$, to make sure that the
output of $T$ on the simulated backward pass is a suffix of $t_2^*$
(the pointer moves backward, coming back to the last position of $t_2$
whenever it reaches the first position of $t_2$). Once
the verification on $t_1$ is done, $T'$ starts, by using a pointer
initially at the first position in $t_2$, to verify that the output of
$T$ in the forward pass is a prefix of $t_2^*$. Once the 
forward and the simulated backward passes merge, the two pointers on
$t_2$ must be at the same position, otherwise the run is
rejected.

 During this verification, $T'$ also has to output a power of
$t_2$ (remind that it has already output $t_1$).  However the
transitions of $T$ may not output exactly one $t_2$, nor a power of
$t_2$, but may cut $t_2$ before its end. Therefore $T'$ needs another
pointer $h$ to know where it is in $t_2$.  
Initially this pointer is at the first position of $t_2$
($h=1$). Suppose that $T'$ simulates $T$ using the (forward)
transition $(p,\sigma,x,p',+1)$ and the (backward) transition
$(q',\sigma',y,q,-1)$.  If this step occurs before the end of $t_1$,
then $T'$ outputs $t_2^\omega[h\dots (h+|y|)]$ ($t_2^\omega$ is the
infinite concatenation of $t_2$), and the pointer $h$ is updated to
$1+ ((h+|y|-1)\ mod\ |t_2|)$. Otherwise,
$T'$ outputs $t_2^\omega[h\dots (h+|x|+|y|)]$ and $h$ is updated to
$1+((h+|x|+|y|-1)\ mod\ |t_2|)$.

The second case (when $T'$ guesses that $t_1$ ends during the backward pass) is similar. $T'$ has to guess
exactly the position in the output where $t_1$ ends. On the first pass it verifies that the output is
a prefix of $t_1$, and on the simulated backward pass, it checks that the output is a suffix of $t_2^*$ (and outputs
as many $t_2$ as necessary, like before), until the end of $t_1$ is guessed to occur. From that moment it enters
a verification mode on both passes.

The main property of this construction is that no wrong output words are produced by $T'$, due
to the verification and the way the output words are produced, i.e. for all $(u,v)\in R(T')$, 
we have $(u,v)\in R(T)$.
\begin{proposition}\label{prop:equivTprime}
    Let $T\in\ZNFT$. $R(T') = \{ (u,v)\in R(T)\ |\ (u,v)\models \pptone\}$.
\end{proposition}

\begin{lemma}\label{lem:EquivBackAndForth}
    Let $T\in \ZNFT$. If $T\models \ppt$, then $T$ is equivalent to the $\eZNFT$ $T'$. 
    Moreover, the latter is decidable.
\end{lemma}

\begin{IEEEproof}
    If $T\models \ppt$, then by Proposition \ref{prop:outputsnecessary}, 
    $T\models \pptone$. Therefore by Proposition \ref{prop:equivTprime}, 
    $T$ and $T'$ are equivalent.

    We know that $R(T') \subseteq R(T)$, and since $T$ and $T'$ are both functional, 
    they are equivalent
    iff $\dom(T)\subseteq \dom(T')$. Both domains can be defined by
    \NFAs. Those \NFAs simulate the three passes in parallel and make sure 
    that those passes define a \zmotion. Therefore testing the equivalence
    of $T$ and $T'$ amounts to test the equivalence of two \NFAs.
\end{IEEEproof}

\subsection{From $\eZNFT$ to $\NFT$}
We have seen how to go from a \ZNFT to an \eZNFT. We now 
briefly sketch how to go from an \eZNFT to a (functional) \NFT.
Given an \eZNFT $T'$, we define an \FUNNFT $T''$ such that 
$T'$ and $T''$ are equivalent as soon as $T'\models\ppt$.
The ideas are very similar to the previous
construction therefore we do not give
all the details here.

\input{figures/property_P2}

We exhibit a property on the form of output words
produced by an \eZNFT that verifies $\ppt$.
 Intuitively, apart from
the beginning of the first pass, and the end of the second pass, if
the two passes produce long enough outputs, then these outputs can be
decomposed so as to exhibit conjugate primitive roots.

\begin{definition}[$\ppttwo$-property]
    Let $T' \in \eZNFT$ with $m$ states, and let $(u,v)\in R(T')$ where 
    $u$ has length $n$. Let $K = 2om^3|\Sigma|$ where $o = max\{|v|\ |\ (p,a,v,q,m)\in \Delta\}$.
    The pair $(u,v)$ satisfies the property $\ppttwo$, denoted by 
    $(u,v)\models \ppttwo$, if for all 
    accepting runs $\rho$ on $u$, 
    there exist two positions $1\leq \ell_1 \leq
  \ell_2 \leq n$ and $w,w',t_1,t_2,t_3\in \Sigma^*$ such that:
$$
\begin{array}{ll}
\out_1[1,\ell_1]   = w                            &  |t_i|\leq 3.K,\  \forall i\in \{1,2,3\}\\
\out_3[\ell_2,n+1] = w'                           &  |\out_1[\ell_2,n]| \leq 3.K \\
\out_1[\ell_1,n]\out_3[1,\ell_2] \in t_1t_2^*t_3   &  |\out_3[1,\ell_1]| \leq 3.K
\end{array}
$$
This decomposition is depicted in Fig.~\ref{fig:ppttwo}.
  $T'$ satisfies property $\ppttwo$, denoted $T\models \ppttwo$, if
  all $(u,v)\in R(T')$ satisfy it. 
\end{definition}

The proof of the following proposition uses the same structure and techniques as 
that of Proposition~\ref{prop:outputsnecessary}. 
Using a (long) case analysis, we identify loops in runs, and apply Property $\ppt$ to show that
output words have the expected form.
\begin{proposition}\label{prop:pptimpliesppttwo}
  Let $T'\in\eZNFT$. If $T'\models\ppt$, then $T'\models\ppttwo$.
\end{proposition}

We can now sketch the construction of an \FUNNFT $T''$ which recognizes
the subrelation of $T'$ defined as $\{(u,v)\in R(T')\mid (u,v)\models
\ppttwo\}$. Again, the construction is rather similar and uses the same techniques 
to that of $T'$ starting from $T$.

The transducer $T''$ simulates, in a single forward pass, the three passes
of $T'$. Hence it also checks that the run of the \ZNFT $T'$ it simulates
is a \zmotion run, which is a semantic restriction of accepting runs of
\ZNFTs. The \FUNNFT $T''$ also guesses positions $\ell_1$ and $\ell_2$,
and uses three modes accordingly.
It also guesses the words $t_1$, $t_2$ and $t_3$, and words for
$\out_3[1,\ell_1]$ and $\out_1[\ell_2,n]$, which are all of bounded length
(see Property~$\ppttwo$).
The output of $T''$ is produced according to the mode, 
using pointers to check the guesses, similarly to $T'$.

If all the guesses happen to be verified, it outputs the correct
output word, otherwise the input word is rejected. As a consequence,
$T''$ recognizes a subrelation of $T'$
and thus checking the equivalence of $T'$ and $T''$ amounts to
checking the equivalence of their domains (as the two transducers are
functional), which is decidable. From
Proposition~\ref{prop:pptimpliesppttwo} we get:

\begin{lemma}\label{lem:EquivTwoForward}
  Let $T'\in \eZNFT$. If $T'\models \ppt$, then $T'$ is equivalent to
  the \FUNNFT $T''$. Moreover, the latter property is decidable.
\end{lemma}

\vspace{1mm}
\noindent \textbf{Proof of Lemma~\ref{lem:characterization}.}
Lemma~\ref{lem:z-cns} states that if $T$ is \NFT-definable, then $T\models\ppt$.
  Conversely, if $T\models \ppt$, then by Lemma \ref{lem:EquivBackAndForth}, the
  first construction outputs an equivalent \eZNFT $T'$. By
  Lemma~\ref{lem:ppt-equiv}, we have $T'\models \ppt$. By Lemma
  \ref{lem:EquivTwoForward}, the second construction outputs an
  equivalent \NFT $T''$. Therefore $T$ is \NFT-definable by $T''$.  In
  order to decide whether $T\models \ppt$, it suffices to construct
  $T'$, check that $T$ and $T'$ are equivalent, and then construct
  $T''$ and check whether $T'$ and $T''$ are equivalent. Both problems
  are decidable by Lemma \ref{lem:EquivBackAndForth} and
  \ref{lem:EquivTwoForward}.

%% file: figures/znft-to-nft.tex
\begin{figure}[t]
\begin{tikzpicture}[->,>=latex,node distance=3.6cm]
 \begin{scope}
 \node (1) {$T\in \ZNFT$};
  \node (2) [right of=1] {$T'\in\eZNFT$};
  \node (3) [right of=2] {$T''\in \FUNNFT$};
  \path 
  (1) edge [bend left,above,pos=.5] node {\small $T\models \ppt \Rightarrow T\equiv T' $} (2)
  (2) edge [bend left,above,pos=.6] node {\small $T'\models \ppt \Rightarrow T'\equiv T''$} (3);
  \end{scope}
  
       \begin{scope}[yshift=-1.1cm,xshift=-.7cm]
       \draw [->,>=latex] 
       (.2,0)    -- (1,0)  .. controls (1.38,0)  and (1.38,.3) ..
       (1,.3)   -- (.5,.3) .. controls (.12,.3) and (.12,.6) ..
       (.5,.6) -- (1.3,.6) ;
    \end{scope}

       \begin{scope}[yshift=-1.1cm,xshift=2.8cm]
       \draw [-] 
       (.2,0)    -- (1,0)  .. controls (1.38,0)  and (1.38,.1) ..  (1.38,.15);
       \draw [-][dotted](1.38,.15) .. controls (1.38, .2) and (1.38, .3) .. 
       (1,.3)   -- (.5,.3) .. controls (.2,.3) and (.2,.45) ..  (.2,.45);
       \draw  [->] (.2,.45) .. controls (.2, .50) and (.2, .6) .. 
       (.5,.6) -- (1.4,.6) ;
       
       \draw[-] (.8,.4) node {\scriptsize $\epsilon$};
       \end{scope}

       \begin{scope}[yshift=-.8cm,xshift=6.4cm]
       \draw [->,>=latex] 
       (0,0)    -- (1.4,0) ;       
       \end{scope}
\end{tikzpicture}
\vspace{-15pt}
\caption{From \ZNFT to \NFT.\label{fig:ZNFT2NFT}}
\vspace{-17pt}
\end{figure}

%% file: figures/property_P1.tex
\usetikzlibrary{decorations.pathreplacing}

\begin{figure}[t]
   \centering
   \scriptsize
   \begin{tikzpicture}[]

     \draw [->,>=latex] 
       (0,0)    -- (5,0)  .. controls (5.7,0)  and (5.7,.8) ..
       (5,.8)   -- (.5,.8) .. controls (-.2,.8) and (-.2,1.6) ..
       (.5,1.6) -- (5.5,1.6) ;

     \draw[densely dotted] (1.8,-.3) -- (1.8,1);

      \draw (0,0) node {$\bullet$};

     \draw (1.8,-.5) node {$\ell$};

    \draw [decorate,decoration={brace,amplitude=5pt}]
(1.7,-0.15) -- (0.2,-0.15) node [black,midway,yshift=-9pt] {$w$};
    \draw [decorate,decoration={brace,amplitude=5pt}]
(5.2,-0.15) -- (1.9,-0.15)  node [black,midway,yshift=-9pt] {$x$};

    \draw [decorate,decoration={brace,amplitude=5pt}]
(0.3,.95) -- (1.7,.95) node [black,midway,yshift=9pt] {$t_3$};
    \draw [decorate,decoration={brace,amplitude=5pt}]
(1.9,.95) -- (5.2,.95) node [black,midway,yshift=9pt] {$y$};

    \draw [decorate,decoration={brace,amplitude=5pt}]
(0.3,1.75) -- (5.2,1.75) node [black,midway,yshift=10pt] {$w'$};

    \draw (3.4,.4) node {$xy\in t_1t_2^*$};


   \end{tikzpicture}
   \vspace{-10pt}
   \caption{Decomposition of the output according to Property $\pptone$.
     \label{fig:pptone}}
   \vspace{-10pt}
\end{figure}

%% file: figures/loop.tex
\usetikzlibrary{decorations.pathreplacing}

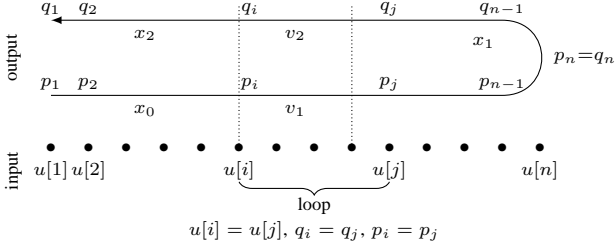
\begin{figure}[t]
   \centering
   \scriptsize
   \begin{tikzpicture}[]

     \foreach \x in {0,.5,...,6.5} \draw (\x,.3) node {$\bullet$};

     \draw (  0,0) node {$u[1]$};
     \draw ( .5,0) node {$u[2]$};
     \draw (2.5,0) node {$u[i]$};
     \draw (4.5,0) node {$u[j]$};
     \draw (6.5,0) node {$u[n]$};

     \draw (-.5,0) node {\rotatebox{90}{input}};
     

     \draw [->,>=latex] 
     (0,1) -- (6,1) .. controls (6.7,1)  and (6.7,2) .. (6,2) -- (0,2);

     \draw (   0,1.15) node {$p_1$};
     \draw (  .5,1.15) node {$p_2$};
     \draw (2.65,1.15) node {$p_i$};
     \draw ( 4.5,1.15) node {$p_j$};
     \draw (   6,1.15) node {$p_{n{-}1}$};

     \draw (   0,2.15) node {$q_1$};
     \draw (  .5,2.15) node {$q_2$};
     \draw (2.65,2.15) node {$q_i$};
     \draw ( 4.5,2.15) node {$q_j$};
     \draw (   6,2.15) node {$q_{n{-}1}$};

     \draw (7.1,1.5) node {$p_n{=}q_n$};

     \draw (-.5,1.5) node {\rotatebox{90}{output}};


     \draw[densely dotted] (2.5,.3) -- (2.5,2.2);    
     \draw[densely dotted] (  4,.3) -- (  4,2.2);    
    
     \draw [decorate,decoration={brace,amplitude=5pt}]
     (4.5,-.15) -- (2.5,-.15) node [black,midway,yshift=-9pt] {loop};

     \draw (3.5,-.8) node {$u[i]=u[j]$, $q_i=q_j$, $p_i=p_j$};

     
     \draw (1.25,0.8) node {$x_0$};
     \draw (3.25,0.8) node {$v_1$};
     \draw (5.75,1.7) node {$x_1$};
     \draw (3.25,1.8) node {$v_2$};
     \draw (1.25,1.8) node {$x_2$};

   \end{tikzpicture}
   \vspace{-10pt}
   \caption{\label{fig:zmotionrun} 
     Decomposition of the two first passes of a \zmotion run with loop.}
   \vspace{-10pt}
\end{figure}

%% file: figures/property_P2.tex
\usetikzlibrary{decorations.pathreplacing}

\begin{figure}[t]
   \centering
   \scriptsize
   \begin{tikzpicture}[]
     
     \draw [-] 
     (0,0) -- (5,0) .. controls (5.7,0)  and (5.7,.4) .. (5.7,.4);
     \draw[-][dotted] 
     (5.7,.4) .. controls (5.7,.4)  and (5.7,.8) .. (5,.8) -- 
     (.5,.8) .. controls (-.2,.8) and (-.2,1.2) .. (-.2,1.2);
     \draw[->,>=latex]
     (-.2,1.2) .. controls (-.2,1.2) and (-.2,1.6) .. (.5,1.6) -- (5.5,1.6);

     \draw (0,0) node {$\bullet$};
     \draw (2.6,.95) node {\scriptsize $\epsilon$};


     \draw[densely dotted] (1.6,-.4) -- (1.6,2);    
     \draw (1.6,-.6) node {$\ell_1$};

     \draw[densely dotted] (3.7,-.4) -- (3.7,2);    
     \draw (3.7,-.6) node {$\ell_2$};
    
 
     \draw [decorate,decoration={brace,amplitude=5pt}]
     (1.5,-0.15) -- (0.2,-0.15) node [black,midway,yshift=-9pt] {$w$};

     \draw [<->]
     (5.2,-0.15) -- (3.8,-0.15)  node [black,midway,yshift=-6pt] {$\le 3K$};
     
     \draw [decorate,decoration={brace,amplitude=5pt}]
     (1.7,.15) -- (5.2,.15) node [black,midway,yshift=9pt] {$v_1$};
     \draw [decorate,decoration={brace,amplitude=5pt}]
     (3.6,1.45) -- (0.3,1.45) node [black,midway,yshift=-9pt] {$v_2$};
     
     \draw [<->] 
     (0.3,1.75) -- (1.5,1.75) node [black,midway,yshift=6pt] {$\le 3K$};

     \draw [decorate,decoration={brace,amplitude=5pt}]
     (3.8,1.75) -- (5.2,1.75) node [black,midway,yshift=10pt,xshift=1pt] {$w'$};

     
     \draw (6.5,1.2) node {$v_1v_2\in t_1t_2^*t_3$};

   \end{tikzpicture}
   \vspace{-10pt}
   \caption{Decomposition of the output according to Property $\ppttwo$.
     \label{fig:ppttwo}}
   \vspace{-10pt}
\end{figure}

%% file: complexity.tex
\section{Discussion}\label{sec:discussion}

\vspace{1mm}
\noindent
{\bf Complexity}
The procedure to decide $(\FUNTW, \NFT)$-definability is 
non-elementary exponential time and space. This is due to the \ZNFT-to-\NFT 
construction which outputs an \NFT of doubly exponential size. 
Indeed, the first step of this construction transforms any \ZNFT
with $n$ states into an \eZNFT with at least $|\Sigma|^{4on^3|\Sigma|}$
states, as the \eZNFT has to guess words of length $4on^3|\Sigma|$, where
$o$ is the maximal length of an output word of a transition. The \eZNFT-to-\NFT construction
also outputs an exponentially bigger transducer. Therefore the 
$\squeezeOp$ operation outputs a transducer which is doubly exponentially
larger. Since this operation has to be iterated $N^2$ times in the worst case,
where $N$ is the number of states of the initial \FUNTW, this leads to a
non-elementary procedure. On the other hand, the best lower bound we have for this problem is
PSpace (by a simple proof that reduces the emptiness problem of the
intersection of $n$ \DFAs is given in Appendix).


\vspace{1mm}
\noindent
{\bf Succinctness}
It is already known that \TWDFAs are exponentially more succinct than
\NFAs \cite{Birget93}. Therefore this result carries over to transducers,  already for transducers defining identity relations on some 
particular domains. However
we show here a stronger result: the succinctness of \TWNFTs also comes 
from the transduction part and not only from the domain part. We can
indeed exhibit a family of \NFT-definable transductions $(R_n)_n$ that can be 
defined by \TWDFTs that are exponentially more succinct than their smallest equivalent
 \NFT, and such that the family of languages $(\dom(R_n))_n$ does not show an exponential blow
up between \TWDFAs and \NFAs.

For all $n\geq 0$, we define $R_n$ whose domain is the set of words $\#u\#$ for all
$u\in \{a,b\}^*$ of length $n$, 
and the transduction is the mirror transduction, i.e. $R_n(\#u\#) = \#\overline{u}\#$.

Clearly, $R_n$ is definable by a \TWDFT with $O(n)$ states that counts up to $n$ the length of the input word by a forward pass, 
and then mirrors it by a backward pass. It is also definable by an \NFT with $O(2^n)$ states: the \NFT guesses a word
$u$ of length $n$ (so it requires $O(2^n)$ states), outputs its reverse, and then verifies that the guess was correct. It is 
easy to prove that any \NFT defining $R_n$ needs at least $2^n$ states by a pumping argument. 
On the other hand, the domain of $R_n$ can be defined by a \DFA with $O(n)$ states that counts the length of the input word up to $n$. Note that the alphabet does not depend on $n$. 



%% file: application.tex
\vspace{1mm}
\noindent
{\bf Further Questions}
We have shown that $(\FUNTW, \NFT)$-definability is decidable, however
with a non-elementary procedure. We would like to characterize precisely
the complexity of this problem. Our procedure works for functional
\TWNFTs, which are equivalent to \TWDFTs. Therefore
we could have done our proof directly for \TWDFTs. However
(functional) non-determinism was added with no cost in the proof so we
rather did it in this more general setting. The extension of our
results to relations instead of functions is still open.


Our proof is an adaptation of the proof of Rabin and Scott \cite{RabinScott59} 
to transducers. Alternative constructions based on the proofs of 
Shepherdson \cite{Shepherdson59} or Vardi \cite{Vardi89}, and
alternative models such as streaming string transducers \cite{alur_et_al:LIPIcs:2010:2853} or
MSO transformations \cite{Cou94,EngHoo01}, could lead to better
complexity results or refined results. In particular, we believe that 
our results are highly related to the problem of minimizing the number
of variables in a streaming string transducer.

Finally, we plan to study extensions of our results to infinite
string tranformations, defined for instance by streaming string
transducers \cite{DBLP:conf/lics/AlurFT12}, and to tree
transformations, following our initial motivation from XML
applications.

{\bf Acknowledgements} We warmly thank Sebastian Maneth and Julien
Tierny for interesting discussions.

%% file: appendix.tex
\section{Complements to Section~\ref{sec:generalcase}}
\label{app:general}

\input{removal}


\section{Complements to Section~\ref{sec:zmotions}}
\label{app:zmotions}

\input{useful}

\input{forward}

\section{Lower Bound}

\begin{lemma}
  (\TWDFT, \NFT)-definability is PSpace-Hard.
\end{lemma}

\begin{IEEEproof}
    Consider $n$ \DFAs $A_1,\dots,A_n$. Let us define the following transduction (where $\#\not\in\Sigma$):
    $$
    \begin{array}{llllllll}
    T & : & u & \mapsto & \left\{ \begin{array}{llll}
            \overline{u_1}\text{ if } u = \#u_1\# u_2\#\text{ and } u_2\in\bigcap_i L(A_i) \\
            \text{undefined otherwise.}
    \end{array}\right.
\end{array}
    $$
    
    Clearly, $T$ is definable by a \TWDFT. It suffices to first perform
    $n$ back and forth non-producing passes on $u$ to determine whether $u_2\in\bigcap_i L(A_i)$, and then
    a last backward pass to reverse $u_1$.

    Then, $T$ is \NFT-definable iff $\dom(T) = \emptyset$ iff $\bigcap_i L(A_i) = \emptyset$. 
    Indeed, if $\dom(T) = \emptyset$ then $T$ is obviously \NFT-definable. Otherwise, there exists
    $u_2\in\bigcap_i L(A_i)$, and therefore $\#\Sigma^*\#u_2\#\subseteq \dom(T)$. If $T$ is \NFT-definable,    then so would be the reverse operation. Contradiction.
\end{IEEEproof}

%% file: removal.tex
\subsection{Iterative \zmotions removal (proof of Lemma~\ref{lem:removalsquare})}
\label{app:removal}

We define the crossing number of the position 
$x \in\{1, \dots, m\}$ as the number $|\{i \mid a_i= x\}|$.
Hence the sequence $s$ is $K$-crossing if all its positions  
$x\in \{1,2\dots, m\}$ have a crossing number less or equal than $K$. 

We say that two \zmotions $z_1=(r_i,r_{i+1})$, $z_2=(r_j,r_{j+1})$ are \emph{consecutive}, resp. \emph{positionally disjoint}, if
$j=i+2$, resp. $\max(a_{r_i},a_{r_{i+1}}) < \min(a_{r_j},a_{r_{j+1}})$ (or  $\max(a_{r_j},a_{r_{j+1}}) < \min(a_{r_i},a_{r_{i+1}})$). 
Moreover we say that $z_1$ and $z_2$ are \emph{disjoint} if they are not consecutive or if they are positionally disjoint.
Equivalently, the \zmotions $z_1 = a_{k_1}, a_{k_1 + 1},\dots, a_{k_2}$ and $z_2= a_{k_3}, a_{k_3 + 1}, \dots, a_{k_4}$ are disjoint 
if and only if $k_2 < k_3$ or $k_4 < k_1$.

\begin{lemma}\label{lemma:consecutive}
If $s$ is $K$-crossing, then for all $z_1, z_2, \dots, z_t$ consecutive \zmotions,
for all $i\leq t-K$, $z_i$ and $z_{i+K}$ are positionally disjoint. 
\end{lemma}
\begin{IEEEproof}
Let $j\in\{1,\dots ,l\}$ such that $z_1=(r_j,r_{j+1}), z_2=(r_{j+2}, r_{j+3}), \dots$ and, wlog, assume $a_{r_j} < a_{r_{j+1}}$. 
As a consequence of the definition of \zmotions,
 consecutive \zmotions  form a stair, that is, we have 
 $a_{r_{j+2i}} \leq a_{r_{j+2(i+1)}}$ and $a_{r_{j+2i+1}} \leq a_{r_{j+2(i+1)+1}}$.
If  $z_i$ and $z_{i+K}$ are not positionally disjoint, all $z_k$
 for $i\leq k \leq i+K$ share the leftmost position of $z_i$, i.e. they share $a_{r_{j+2(i-1)}}$.
 Therefore $s$ is not $K$-crossing.
\end{IEEEproof}

We say that a position $x$ is \emph{in between} the positions $y$ and $z$ whenever $y \leq x \leq z$ or $z \leq x \leq y$.
We say that the pair of reversals (or a \zmotion) $(r,s)$ is \emph{nested} 
into the pair $(r',s')$ if $a_r$ and $a_s$ are in between $a_{r'}$ and $a_{s'}$.

\begin{lemma}\label{lemma:nesting}
Let $(r_i,r_j)$, with $i<j$, be a pair of reversals and $z=(r,r')$ be a \zmotion.
 If $a_{r}$ is in between $a_{r_{i}}$ and $a_{r_{j}}$, and $r\in\{r_{i+1}, r_{j+1}\}$,
 or if $a_{r'}$ is in between $a_{r_{i}}$ and $a_{r_{j}}$, and $r'\in\{r_{i-1}, r_{j-1}\}$,
then $z$ is nested in $(r_i,r_j)$.
\end{lemma}
\begin{IEEEproof}
Suppose that $z=(r_{i-2}, r_{i-1})$ (the other cases are proved similarly).
Wlog assume $a_{r_{j}} \leq a_{r_{i}}$, so by hypothesis we have $a_{r_{j}} \leq a_{r_{i-1}} \leq a_{r_{i}}$.
Then, as a consequence of basic properties of reversals, 
$a_{r_{i-1}} \leq a_{r_{i-2}}$ (because $a_{r_{i-1}} \leq a_{r_{i}}$).
Moreover as $(r_{i-2}, r_{i-1})$ is a \zmotion we have  $a_{r_{i-2}} \leq a_{r_{i}}$.
Therefore we have the inequalities: $a_{r_{j}} \leq a_{r_{i-1}} \leq a_{r_{i-2}} \leq a_{r_{i}}$, which means that $z$ is nested in $(r_i,r_j)$.
\end{IEEEproof}

The one-step sequence $s'$ is obtained from $s=a_1,\dots, a_n$ by removing the \zmotion $z = a_{k_1},\dots, a_{k_2}$
also defined by its reversals as $z=(r_i,r_{i+1})$, if $s'=a_1, \dots, a_{r_i}, a_{k_2+1}, \dots, a_n$. 
Note that the sequence $s'$ is a one-step sequence because $s$ is one and because $a_{r_i} = a_{k_2}$. 
The sequence $s'$ has exactly 2 less reversals than $s$ and 
each reversal of $s$ not in $z$ corresponds to one of the reversals of $s'$, 
each \zmotion $z'=(r_j,r_{j+1})$ of $s$ such that $r_j, r_{j+1}\notin \{r_i, r_{i+1}\}$ is also a \zmotion in $s'$ (up to an index shift). 
Note also that positionally disjoint \zmotions in $s$ are still positionally disjoint in $s'$.

We define the function $\squeeze s$ as the function that associates to a one-step sequence $s$ the 
set of one-step sequences that can be obtained from $s$ by removing \emph{some} pairwise disjoint \zmotions of $s$.

We say that a set $Z$ of \zmotions of $s$ is \emph{consistent} if no two \zmotions of $Z$ share a reversal,
that is, if $(r,r'), (s,s') \in Z$, then $r,r'\neq s$ and $r,r'\neq s'$.  
The consistent set $Z$ is \emph{maximal} if it is not strictly contained into any other consistent set  of \zmotions of $s$.
\begin{lemma}\label{lemma:stairsremoval}
Let $s$ be a $K$-crossing one-step sequence.
If $Z$ is a consistent set of \zmotions of $s$ then 
there is some $s'\in \squeezepow{K}{s}$ that contains no \zmotion of $Z$.
\end{lemma}
\begin{IEEEproof}
Let $Z=\{z_1, z_2, \dots\}$, where the \zmotion are ordered, i.e., if $z_i=(r,r+1)$ and $z_{i+1}=(s,s+1)$ then $r+1<s$.
We define $s_0=s$, and, for all $0< i \leq K$, $s_{i}$ is obtained from $s_{i-1}$ by removing $z_{i}, z_{i+K}, z_{i+2K}\dots$. 
Clearly $s_{i+1}\in \squeezepow{i+1}{s}$ if $ z_{i+j K}$ and $z_{i+j' K}$ are disjoint in $s_i$.
We consider two cases.
Either  $ z_{i+j K}$ and $z_{i+j' K}$ belong to a sequence of \zmotions in $Z$ that are consecutive in $s$, in that case we can apply Lemma~\ref{lemma:consecutive} which shows that they are disjoint. 
Otherwise, $ z_{i+j K}$ and $z_{i+j' K}$ do not belong to such a sequence of \zmotions in $Z$, that is, 
there exists a reversal $r$ that does not appear in any \zmotion of $Z$ and 
which is between the second reversal of $ z_{i+j K}$ and the first reversal of $z_{i+j' K}$,
but then they cannot be consecutive in $s'$ (they are also separated by $r$ in $s'$), so, by definition, they are disjoint.
\end{IEEEproof}

\begin{IEEEproof}[Proof of Lemma~\ref{lem:removalsquare}]
Let $s_1=s$, and for all $i\geq 1$, let $Z_i$ be a maximal consistent set of \zmotions of $s_i$, 
and $s_{i+1}$ be the one-step sequence obtained from $s_i$ by removing $Z_i$.
We show that each \zmotion $z$ in $Z_i$ 
has one of its positions whose crossing in $s$ is at least $i+i'$ where $i'$ is the crossing of the corresponding (some shift might be applied) position in $s_{i+1}$.
This trivially holds for $s_1=s$, so suppose it holds for $i$ and let us show it also holds for $i+1$.
Let $r'_1, r'_2, \dots, r'_{l'}$ be the reversals of $s_i$, let  $z=(r'_{i'},r'_{j'})$ be a \zmotion in $Z_{i+1}$ 
(recall that we abuse notation and refer to the reversals of $s_{i+1}$ using the reversals of $s_i$ though there is a shift of index for some of them).
As $Z_i$ is maximal, $z$ is not a \zmotion in $s_i$, so there is a \zmotion $z'=(r'_k, r'_{k+1})\in Z_i$ such that one of the following holds:
\begin{itemize}
\item $k=i'+1$ or $k=j'+1$ and $a_{r'_k}$ is in between $a_{r'_{i'}}$ and $a_{r'_{j'}}$
\item $k=i'-2$ and $a_{r'_{k+1}}$ is in between $a_{r'_{i'}}$ and $a_{r'_{j'}}$
\end{itemize}
Intuitively the above property states that one of the \zmotions, $z'$, in $Z_i$ must prevent $z$ to be a \zmotion in $s_i$, that is,
$z'$ is somehow 'in' $z$.   
In each of these two cases we can apply  Lemma~\ref{lemma:nesting} which states that $z'$ is nested in $z$. 
By induction hypothesis, one of the position of $z'$ has a crossing number in $s$ of at least $i+i'$, where $i'$ is the crossing number of the corresponding position in $s_{i+1}$.
As $s_{i+2}$ is obtained from $s_{i+1}$ after removing $z$, we have $i' \geq 1+ i''$ where $i''$ is the crossing number of the corresponding position in $s_{i+2}$. 
So we have proved that the crossing number of this position is at least $(i+1) + i''$.

To conclude, as $s$ is $K$-crossing, all positions are at most $K$-crossing, therefore the property we just proved implies that $s_i$ for $i>K$ has no \zmotion, that is $s_i=1,2,\dots, m$. By Lemma~\ref{lemma:stairsremoval}, 
$K$ applications of $\squeezeOp$ are sufficient to remove a consistent set of \zmotions, therefore $1,2, \dots, m$ is in $\squeezepow{K^2}{s}$.
\end{IEEEproof}

\input{app-gene}

\input{NC}

\subsection{Definition of $\squeeze{T}$}

We let $L_T(q_1,q_2)  =  (Q^{q_1,q_2}, q_0^{q_1,q_2}, F^{q_1,q_2},\Delta^{q_1,q_2})$
and $\overline{R_T(q_1,q_2)} = (P^{q_1,q_2}, p_0^{q_1,q_2}, G^{q_1,q_2},\Gamma^{q_1,q_2})$ 
for all $q_1,q_2\in Q$.

We let $\squeeze{T} = (Q', Q'_0, F', \Delta')$  and show formally how to construct it. For more convenience
here we assume that $\squeeze{T}$ can have a set of initial states. It will be easy to transform it into a (usual) \TWNFT. 
We let $Q' = Q\uplus \biguplus   \{Q^{q_1,q_2}\uplus P^{q_1,q_2}\ |\ q_1,q_2\in Q\}$, $Q'_0 = \{q_0\}\cup \{q_0^{q_1,q_2}\ |\ q_1,q_2\in Q\}\cup \{p_0^{q_1,q_2}\ |\ q_1,q_2\in Q\}$, $F' = F$ and $\Delta'$ is the least set satisfying for all $q_1,q_2\in Q$:

\begin{itemize}
\item $\Delta \uplus \biguplus_{q_1,q_2\in Q} \Delta^{q_1,q_2}\subseteq \Delta'$;
\item $\forall (p,a,v,q_1,m){\in} \Delta$, $(p,a,v,q_0^{q_1,q_2},m){\in} \Delta'$;
\item $\forall q\in F^{q_1,q_2}$, $\forall (p,a,v,q,+1)\in \Delta^{q_1,q_2}$, 
  $\forall (q_2,a,v',q_3,m)\in \Delta$, $(p,a,vv',q_3,m)\in\Delta'$

\item $\forall q\in F^{q_1,q_2}$, $\forall (p,a,v,q,+1)\in \Delta^{q_1,q_2}$, 
  $\forall (q_2,a,v',q_3,m)\in \Delta$, for all $q_4\in Q$, 
  $(p,a,vv',q_0^{q_3, q_4},m)\in\Delta'$

\item $\forall q\in F^{q_1,q_2}$, $\forall (p,a,v,q,+1)\in \Delta^{q_1,q_2}$, 
  $\forall q_3\in Q$,
  $\forall (q_0^{q_2,q_3},a,v',q',m)\in \Delta^{q_2,q_3}\cup \Gamma^{q_2,q_3}$, $(p,a,vv',q',m)\in\Delta'$
\end{itemize}

and similarly:

\begin{itemize}
\item $\biguplus_{q_1,q_2\in Q} \Gamma^{q_1,q_2}\subseteq \Delta'$;
\item $\forall (p,a,v,q_1,m){\in} \Delta$, $(p,a,v,p_0^{q_1,q_2},m){\in} \Delta'$;
\item $\forall q\in G^{q_1,q_2}$, $\forall (p,a,v,q,-1)\in \Gamma^{q_1,q_2}$, 
  $\forall (q_2,a,v',q_3,m)\in \Delta$, $(p,a,vv',q_3,m)\in\Delta'$

\item $\forall q\in G^{q_1,q_2}$, $\forall (p,a,v,q,-1)\in \Gamma^{q_1,q_2}$, 
  $\forall (q_2,a,v',q_3,m)\in \Delta$, for all $q_4\in Q$, 
  $(p,a,vv',q_0^{q_3, q_4},m)\in\Delta'$

\item $\forall q\in G^{q_1,q_2}$, $\forall (p,a,v,q,-1)\in \Gamma^{q_1,q_2}$, 
  $\forall q_3\in Q$,
  $\forall (q_0^{q_2,q_3},a,v',q',m)\in \Delta^{q_2,q_3}\cup \Gamma^{q_2,q_3}$, $(p,a,vv',q',m)\in\Delta'$
\end{itemize}

\subsection{Proof of Proposition \ref{prop:correctness}}

\begin{IEEEproof}
    Since $\squeeze{T}$ contains $T$ as a subtransducer, we have
    $R(T)\subseteq R(\squeeze{T})$. Let us show that
    $ R(\squeeze{T}) \subseteq R(T)$. Let $(u,v)\in R(\squeeze{T})$. 
    Therefore there exists an accepting run $\rho$ of $\squeeze{T}$ on $u$
    that outputs $v$. We are going to construct an accepting run of $T$ on $u$
    that outputs $v$, this can be done by induction on the number of times $\rho$
    goes in \Zmode. If it never does so, $\rho$ is accepting run of $T$ and
    we are done. Otherwise suppose that $\rho$ goes at least once in
    \Zmode for some $q_1,q_2\in Q$. Note that the set $\Delta'$ consists
    of $\Delta$, the sets $\Delta^{p,q}$ and $\Gamma^{p,q}$ for all $p,q\in Q$, and
    new transitions of three kinds (of the form $(p,a,vv',q_3,m), (p,a,vv',q_0^{q_3,q_4},m)$ and
    $(p,a,vv',q',m)$ in the definition). Consider the first use of such a transition $t$ in $\rho$.
    One can decompose $\rho$ as $\rho_1\rho_2 t \rho_3$ where $\rho_1$ is in \Tmode, 
    $\rho_2$ in \Zmode, and assume that $\rho_2t$ is a forward run on a factor $u_2$ of $u$
    (the case of a backward run is symmetric).

    Let us inspect the case where 
    $t = (p,a,vv',q_3,m)$. The other two cases (depending on the form of $t$) are 
    proved similarly. Suppose that $p\in Q^{q_1,q_2}$. Then it means that $(u_2,v)\in \lc_T(q_1,q_2)$,
    and therefore one can easily reconstruct a \zmotion run $\rho'_2$ of $T$ on $u_2$ from $q_1$ to $q_2$
    that outputs $v$. Then by definition of $\Delta'$, we know that there exists a transition
    from $q_2$ to $q_3$ that produces $v'$. By induction we can also transform
    $\rho_3$ into a run $\rho'_3$ of $T$ that ends in an accepting state and outputs the same word. 
    Therefore $\rho'_1\rho'_2(q_2,a,v',q_3,m)\rho'_3$ is an accepting run of $T$ on $u$ that outputs 
    the same word as $\rho$. Therefore $(u,v)\in R(T)$. 
\end{IEEEproof}

%% file: app-gene.tex
\subsection{Proof of Proposition \ref{prop:localzmotions}}

A crossing sequence $s$ is \emph{repetition-free} if each state occurs at most
once in $s$. If $Q$ is the set of states of $A$, we denote by $CS(Q)$ the set of 
repetition-free crossing sequences of  $A$.

Based on Shepherdson's construction, it is possible to construct a one-way
automaton whose states are sequences of states, such that any run $\rho$ of $A$ maps to
the sequence of crossing sequences of $\rho$, and conversely any sequence of crossing sequences
of this automaton maps to a run of $A$. This automaton may have infinitely many states, but it
is well-known it is sufficient to consider repetition-free crossing sequences of states only \cite{hopcroft-ullman:1979a}.

\begin{lemma}[\cite{hopcroft-ullman:1979a}]\label{lem:csconstruction}
    For all \TWNFAs $A$ with set of states $Q$, it is possible to construct 
    an equivalent \NFA $CS(A)$ whose set of states is
    $CS(Q)$, and such that for all accepting runs $\rho'$ of 
    $CS(A)$ on $u$, there exists an accepting run $\rho$ of 
    $A$ on $u$ such that $CS(\rho) = \rho'$.
\end{lemma}

\begin{lemma}\label{lem:factor}
    Let $A$ be a \TWNFA with set of states $Q$, and $q_1,q_2\in Q$. 
    Let $M_{q_1,q_2}$ be the language of words $u_2$ such that there exists a
word $u\in L(A)$, an accepting run $\rho$ of $A$ on
$u$ such that $\rho = \rho_1(q_1,i_1)\rho_2(q_2,i_2)\rho_3$ and
$u_2 = u[i_1..i_2]$. Then $M_{q_1,q_2}$ is regular.
\end{lemma}

\begin{IEEEproof}
Given two sequences of states $s_1$ and $s_2$, the language $Acc_{s_1,s_2}$ is defined 
as the set of words $u_2\in\Sigma^*$ such that there exist a word $u\in\Sigma^*$, two positions 
$i_1\leq i_2$ such that $u_2 = u[i_1..i_2]$, and an accepting run $\rho$ on $u$ such that
$CS(\rho, i_1) = s_1$ and $CS(\rho,i_2) = s_2$. In other words, $s_2$ is accessible
from $s_1$ by $u_2$. It is easy to show that for all $s_1,s_2$, there exists repetition-free
sequences $s'_1,s'_2$ such that $Acc_{s_1,s_2} = Acc_{s'_1,s'_2}$. Therefore one can consider
repetition-free sequences only. We have seen (Lemma \ref{lem:csconstruction}) that one can construct
an \NFA whose states are the repetition-free crossing sequences of the runs of $T$. 
An easy reachability analysis of this \NFA allows one to construct an \NFA $A_{q_1,q_2}$ whose
states are repetition-free crossing sequences of $T$ and such that 
$M_{q_1,q_2} = \bigcup \{ Acc_{s_1,s_2} \ |\ q_1\in s_1,q_2\in s_2,\ s_1,s_2\text{ are repetition-free}\}$.
\end{IEEEproof}

\begin{IEEEproof}
 The transduction $\lc_T(q_1,q_2)$ is a function, otherwise
$T$ would not be functional.

We define an intermediate \ZNFT $Z'_T(q_1,q_2)$ that mi\-mics $T$ but starts initially in the state $q_1$
and whenever it reaches the state $q_2$, it non-deterministically decides to go to a fresh accepting state $q'_f$. 
Formally, $Z'_T(q_1,q_2) = (Q\cup \{q'_f\}, q_1, \{q'_f\}, \Delta')$ where $\Delta' = \Delta\cup \{ (q_2, a, \epsilon, q'_f, +1)\ |\ a\in\Sigma\}$.
Clearly, to any accepting run of $Z'_T(q_1,q_2)$ on a word $u_2\in \Sigma^*$ corresponds a \zmotion run of $T$ on $u_2$ of the form
$\rho'_2 = (q_1,1)\rho_2(q_2,|u_2|)$ and conversely.
However $Z'_T(q_1,q_2)$ is too permissive as it does not check that $\rho'_2$ can
be embedded into a global accepting run of $T$.  We now show how to restrict the domain of $Z'_T(q_1,q_2)$ to
take this further constraint into account.

By a simple adaptation of Shepherdson's construction (see Lemma \ref{lem:factor}), the language
$M_{q_1,q_2}$ of words $u_2$ such that there exists $u\in \dom(T)$ and an accepting run $\rho$ of $T$ on
$u$ such that $\rho = \rho_1(q_1,i_1)\rho_2(q_2,i_2)\rho_3$ and
$u_2 = u[i_1..i_2]$, can be defined by an \NFA $A_{q_1,q_2}$.%
%
The transducer $Z_T(q_1,q_2)$ is finally defined as $Z'_T(q_1,q_2)$ where during the third and last pass, it also checks that the input word is in $M_{q_1,q_2}$ by running 
$A_{q_1,q_2}$ in parallel via a product construction.

Let us briefly explain why this construction is correct. 
Suppose that $(u_2,v)\in Z_T(q_1,q_2)$. We have $u_2\in M_{q_1,q_2}$, therefore there exist
$u\in \Sigma^*$ and two positions $i_1<i_2$ such that $u_2 = u[i_1..i_2]$, and
an accepting run $\rho$ of $T$ of the form $\rho_1(q_1,i_1)\rho_2(q_2,i_2)\rho_3$.
The subrun $(q_1,i_1)\rho_2(q_2,i_2)$ is not necessarily a \zmotion, and
it does not necessarily outputs $v$. However since $(u_2,v)\in Z_T(q_1,q_2)$, we also
have that $(u_2,v)\in Z'_T(q_1,q_2)$, and therefore there exists a \zmotion run $\rho'$ of $T$ from
$q_1$ to $q_2$ on $u_2$. One can therefore substitute $(q_1,i_1)\rho_2(q_2,i_2)$ by $\rho'$ in $\rho$ (modulo a
shift of  the positions occurring in $\rho'$), and one
gets a new run $\gamma = \rho_1\rho'\rho_2$. The run $\gamma$ is still an accepting run of 
$T$ on $u$, and therefore $(u_2,v)\in L_T(q_1,q_2)$. The converse is easy by applying the definitions.
\end{IEEEproof}

%% file: NC.tex
\subsection{Proof of Lemma~\ref{lem:gen-cns}}

\begin{IEEEproof}
  As in the proof of Proposition \ref{prop:localzmotions}, we consider
  only the transductions $\lc_T(q_1,q_2)$, the other case being solved
  by using the equality $\rc_T(q_1,q_2) =
  \lc_{\overline{T}}(q_1,q_2)$.  Let $Z\in\ZNFT$ that defines
  $\lc_T(q_1,q_2)$ for some $q_1,q_2$ and suppose that $T$ is
  \NFT-definable.  By Lemma~\ref{lem:characterization} we have to
  show that $Z\models \ppt$.  Let $u\in \dom(Z)$ of length $n$ and
  $\rho= (p_1,1)\dots (p_n,n)(q_{n-1},n-1)\dots (q_1,1)(r_2,2)\dots
  (r_{n+1},n+1)$ an accepting run of $Z$ on $u$. Let $(i_1,j_1)$ and
  $(i_2,j_2)$ be two loops of $\rho$ such that $j_1\leq i_2$. These
  loops induce a decomposition of the input word $u$ as
  $u=u_1u_2u_3u_4u_5$ with $u_2=u[i_1..j_1-1]$ and
  $u_4=u[i_2..j_2-1]$.

  As $(i_1,j_1)$ and $(i_2,j_2)$ are loops in $\rho$, for any
  $k_1,k_2\geq 0$, we have $u_1u_2^{k_1}u_3u_4^{k_2}u_5\in \dom(Z)$.
  By definition of the transduction $\lc_T(q_1,q_2)$, any word in
  $\dom(Z)$ can be extended into a word in $\dom(T)$. Thus, for any
  $k_1,k_2\geq 0$, there exists
  $\alpha_{k_1,k_2},\alpha'_{k_1,k_2}\in\Sigma^*$ such that
  $u(k_1,k_2) = \alpha_{k_1,k_2}u_1u_2^{k_1}u_3u_4^{k_2}u_5\alpha'_{k_1,k_2}\in
  \dom(T)$. 

  In addition, by assumption, $T$ is \NFT-definable and thus there
  exists an \NFT $T'$ such that $T\equiv T'$. We consider such an \NFT
  $T'$, and denote by $N$ its number of states. Let us consider
  $k_1=k_2=N+1$. There exists an accepting run $\rho'$ of $T'$ on the
  word $u(k_1,k_2)$. Consider the state in which is this run just
  before the $i$-th iteration of the word $u_2$, for $i\in
  \{1,\ldots,k_1\}$. As $k_1=N+1$, two of these states must be
  equal. A similar reasoning can be done for the powers of the word
  $u_4$. As a consequence, there exist constants $c_i,c'_i \geq 0$
  with $i\in\{1,2,3\}$ such that $c_2,c'_2>0$ and the word
  $u(k_1,k_2)$ can be decomposed as follows:
  $$
  u(k_1,k_2) = \alpha_{k_1,k_2} u_1 u_2^{c_1} u_2^{c_2} u_2^{c_3} u_3
  u_4^{c'_1} u_4^{c'_2} u_4^{c'_3} u_5 \alpha'_{k_1,k_2}
  $$
  with the property that $\rho'$ contains two loops on the input
  subwords $u_2^{c_2}$ and $u_4^{c'_2}$.
  
  To conclude, we let $\beta_1$ (resp. $\beta_2$, $\beta_3$,
  $\beta_4$, $\beta_5$) be the output produced by $\rho'$ on the input
  subword $ u_1 u_2^{c_1}$ (resp. $u_2^{c_2}$,
  $u_2^{c_3}u_3u_4^{c'_1}$, $u_4^{c'_2}$, $u_4^{c'_3}u_5$), and
  $f(k_1,k_2)$ (resp. $g(k_1,k_2)$) be the output produced by $\rho'$
  on the input subword $\alpha(k_1,k_2)$ (resp. $\alpha'_{k_1,k_2}$).

\end{IEEEproof}

%% file: useful.tex
\subsection{Technical results}

\begin{lemma}\label{lem:pumpingfat}
Let $\Sigma,\Gamma,\Lambda$ be three finite alphabets, $\Psi$ a morphism 
from $\Gamma$ to $\Sigma^*$ and $\Phi$ a morphism from $\Gamma$ to
$\Lambda$. Let $M = max \{ |\Psi(\gamma)|\ |\ \gamma\in\Gamma\}$. 
For all words $u\in\Gamma^*$, if $|\Psi(u)|>(|\Lambda|+1).M$, then there exist two positions
$1\leq k_1<k_2\leq |u|$ such that\footnote{In this Lemma, if $k_1=1$ then we let $u[1..(k_1-1)] = \epsilon$}:
\begin{enumerate}
    \item $|\Psi(u[1..(k_1-1)])|\leq (|\Lambda|+1).M$
    \item $1\leq |\Psi(u[k_1..(k_2-1)])|\leq (|\Lambda|+1).M$
    \item $\Phi(u[k_1]) = \Phi(u[k_2])$. 
\end{enumerate}
\end{lemma}

\begin{IEEEproof}
    Let $L(u)$ be the set of \emph{loops} that are strictly contained in $u$, i.e.
    $L(u) = \{ (i,j)\ |\ 1\leq i<j\leq |u|,\ (i\neq 1)\vee (j\neq |u|),\ \Phi(u[i]) = \Phi(u[j])\}$. 
    We first show the following by induction on $|u|$:
    $$
    (i)\left\{\begin{array}{c}
            |\Psi(u)|>(|\Lambda|+1).M \\
    \implies \\
    \exists (i,j)\in L(u),\ 
    1\leq |\Psi(u[i..j])|\leq (|\Lambda|+1).M
  \end{array}\right.
    $$ 
    If $|u| = 0$ (resp. $|u|=1$) then $|\Psi(u)|=0$ (resp. $|\Psi(u)\leq M$) and therefore the above implication is obviously satisfied. 
    Otherwise suppose that $|u|>0$ and $|\Psi(u)|>(|\Lambda|+1).M$. Therefore
    we have $|u|>|\Lambda|+1\geq 2$, and
    $|u[2..|u|]|>|\Lambda|$, and
    so by the pigeon-hole principle there exist two positions $i<j$ in $u[2..|u|]$ such that $\Phi(u[i]) = \Phi(u[j])$, 
    so that $L(u)\neq \varnothing$.

    Suppose that for all $(i,j)\in L(u)$, $\Psi(u[i..(j-1)])=\epsilon$. If we remove
    maximally from $u$ all the factors of $u$ from position $i$ to position $(j-1)$
    for all $(i,j)\in L(u)$, 
    one obtains a word $v$ such that $L(v) = \varnothing$ and $|\Psi(v)|=|\Psi(u)|>(|\Lambda|+1).M$. Moreover
    $|v|\leq |\Lambda|+1$ since $L(v) = \varnothing$, but this contradicts $|\Psi(v)|>(|\Lambda|+1).M$
    by definition of $M$. Since $L(u)\neq \varnothing$, we get the existence of
    $(i_0,j_0)\in L(u)$ such that 
    $\Psi(u[i_0..(j_0-1)])\neq \epsilon$. If $|\Psi(u[i_0..(j_0-1)])| \leq (|\Lambda|+1).M$ we are done.
    Otherwise, since $|u[i_0..(j_0-1)]|<|u|$, by induction hypothesis we get the existence of a pair
    $(i^*,j^*)\in L(u[i_0..(j_0-1)])$ 
    such that $1\leq |\Psi(u[i_0..(j_0-1)][i^*..(j^*-1)])|\leq (|\Lambda|+1).M$, from which we can conclude
    by taking $i = i^*+i_0-1$ and $j = j^*+i_0-1$ (note that $(i, j)\in L(u)$).

    This shows items $(2)$ and $(3)$ of the Lemma. Again by induction
    on $|u|$ and by using $(i)$, we prove the lemma. If $|u|=0$ or
    $|u|=1$, then the implication obviously holds. Otherwise assume
    that $|\Psi(u)|>(|\Lambda|+1).M$.  By $(i)$ there exists
    $(k_1,k_2)\in L(u)$ that satisfies $(2)$ and $(3)$. If
    $|\Psi(u[1..(k_1-1)])|\leq (|\Lambda|+1).M$ we are done, otherwise
    by induction hypothesis, there exists $(k'_1,k'_2)\in
    L(u[1..(k_1-1)])$ which satisfies $(1)$, $(2)$ and $(3)$, from
    which we can conclude.
\end{IEEEproof}

\begin{lemma}\label{lem:simpleloopsruns}
  Let $T\in \ZNFT$ with $m$ states. Let $o$ be the maximal length of an
  output word in a transition of $T$ and $K=2.o.m^3.|\Sigma|$. Let
  $\rho$ be a run on a word $u$ of length $n$.  We write $\rho$ as the
  sequence $(p_1,1)\dots (p_n,n)(q_{n-1},n-1)\dots (q_1,1)(r_2,2)\dots
  (r_{n+1},n+1)$ and let $q_n=p_n$ and $r_1=q_1$.
  Let $1\leq k < \ell \leq n$ such that
  $|\out_2[k,\ell]| > K$.
  There exists a loop $(i,j)$ in $\rho$ such that $k\leq i < j \leq \ell$ and
  \begin{enumerate}
    \item $|\out_2[k,i]|\leq K$
    \item $1\leq |\out_2[i,j]|\leq K$.
  \end{enumerate}
\end{lemma}

\begin{IEEEproof}
    We show this result by using Lemma \ref{lem:pumpingfat}.

    We consider the alphabet $\Delta^3\times \Sigma$, where $\Delta$
    denotes the set of transitions of $T$. Given a triple of
    transitions $\theta = ((s_\ell,a_\ell,u_\ell,s'_\ell)_{1\leq \ell
      \leq 3})$, and a letter $a\in \Sigma$, we define the mappings
    $\Psi$ and $\Phi$ as $\Psi(\theta,a) = u_2 $ and $\Phi(\theta,a) =
    (s_1,s_2,s_3,a)$.  Then, we associate to the run $\rho$,
    considered between positions $k$ and $\ell$, a word over this
    alphabet of length $\ell-k$, indexed from $k$ to $\ell-1$, and
    defined as $\eta=(\sigma_m)_{k\leq m \leq \ell-1}$, where
    $\sigma_m$ is composed of the three transitions used respectively
    to go from configuration $(p_m,m)$ to configuration
    $(p_{m+1},m+1)$, from configuration $(q_{m+1},m+1)$ to
    configuration $(q_m,m)$, and from configuration $(r_m,m)$ to
    configuration $(r_{m+1},m+1)$, and of the letter $u[m]$.

    Using these definitions, we have $\Psi(\eta) = \out_2[k,\ell]$,
    and, for any $k\leq m \leq \ell-1$,
    $\Phi(\sigma_k)=(p_k,q_k,r_k,u[k])$.
    Then it suffices to apply Lemma \ref{lem:pumpingfat} to get the result. 
\end{IEEEproof}

\begin{lemma}\label{lem:outputform}
    Let $x,y,z,t\in \Sigma^*$  such that 
    $x\neq \epsilon$ and $y\neq \epsilon$. 
    Suppose that for all $i\geq 0$, 
    $x^i y z^i$ is a prefix of $t^\omega$. Then
    there exists $\alpha_1,\alpha_2\in \Sigma^*$ such that
    $x \in (\alpha_1\alpha_2)^*$, $z\in (\alpha_2\alpha_1)^*$ and
    $xyz\in \alpha_1(\alpha_2\alpha_1)^*$.
\end{lemma}
\begin{IEEEproof}
By Lemma~\ref{lem:fundamental} $\primroot(x)\sim\primroot(t)$ and $\primroot{(z)}\sim\primroot{(t)}$, 
therefore  $\primroot{(x)}\sim\primroot{(z)}$, i.e. there exists $\alpha_1, \alpha_2$
with $x\in (\alpha_1\alpha_2)^*$ and $z\in (\alpha_2\alpha_1)^*$. 
Moreover as $x^i$ is a prefix of $t^\omega$ for all $i>0$, clearly $\primroot{(t)}=\primroot(x)=\alpha_1\alpha_2$.

Now let $xyz= (\alpha_1\alpha_2)^k\alpha$ a prefix of $(\alpha_1\alpha_2)^\omega$ and let us show that $\alpha =\alpha_1$.
So suppose $\alpha\beta=\alpha_1$ (the other case when $\alpha_1\beta = \alpha$ is proved similarly).
Therefore $z=(\alpha_2\alpha_1)^a=(\alpha_2\alpha\beta)^a$ but also $xyz= (\alpha_1\alpha_2)^k\alpha$ implies that 
$z =  (\beta\alpha_2\alpha)^a$. So $\beta\alpha_2\alpha = \alpha_2\alpha\beta$ which means $\alpha_1\alpha_2$ is not primitive if $\beta\neq\epsilon$.
\end{IEEEproof}

\subsection{Proof of Proposition~\ref{prop:outputsnecessary}}

\begin{IEEEproof}
  \noindent $\bullet$ If $|\out_2[1,n-1]| \leq K$, then clearly, it
  suffices to take $\ell = n$, $t_1 = \out_2[n-1,n]$,
  $t_2=\varepsilon$, $t_3 = \out_2[0,n-1]$, $w = \out_1[1,n]$ and $w'
  = \out_3[1,n+1]$.


\noindent $\bullet$ Otherwise, $|\out_2[1,n-1]|>K$. Therefore $u$ is
of length $2.m^3.|\Sigma|$ at least and there exists necessarily a
(non-empty) loop $(i,j)$ in $\rho$.  We can always choose this loop
such that $|\out_2[1,i]| \leq K$ and $1\leq |\out_2[i,j]|\leq K$ (see
Lemma \ref{lem:simpleloopsruns}).

The loop partitions the input and output words into factors that are
depicted in Fig. \ref{fig:zmotionrun} (only the two first passes are
depicted). Formally, let $u = u_1u_2u_3$ such that $u_2 =
u[i..(j{-}1)]$.  Let $x_0 = \out_1[1,i]$, $v_1 = \out_1[i,j]$, $x_1 =
\out_1[j,n]\out_2[j,n]$, $v_2 = \out_2[i,j]$, $x_2 = \out_1[1,i]$,
$x_3 = \out_3[1,i]$, $v_3 = \out_3[i,j]$ and $x_4 = \out_3[j,n+1]$. In
particular, we have $|x_2|\leq K$, $1\leq |v_2| \leq K$ and
$x_0v_1x_1v_2x_2x_3v_4x_4\in T(u)$. Since $(i,j)$ is a loop we also
get $x_0v_1^kx_1v_2^kx_2x_3v_3^kx_4\in T(u_1u_2^ku_3)$ for all $k\geq
0$.

We then distinguish two cases:
\begin{enumerate}
\item If $v_1 \neq \epsilon$. We can apply Property $\ppt$ by taking
the second loop empty. We get that
  for all $k\geq 0$
  $$
  f(k)x_0v_1^{kc+c'}x_1v_2^{kc+c'}x_2x_3v_3^{kc+c'}x_4g(k) =
  \beta_1\beta_2^k\beta_3
  $$
  where $f,g:\mathbb{N}\rightarrow \Sigma^*$, $c\in\mathbb{N}_{>0}$,
  $c'\in\mathbb{N}$, and $\beta_1,\beta_2,\beta_3\in\Sigma^*$.
  Since the above equality holds for all $k\geq 0$, we can apply Lemma
  \ref{lem:fundamental} and we get $\primroot(v_1)\sim
  \primroot(\beta_2)$ and $\primroot(\beta_2) \sim \primroot(v_2)$,
  and therefore $\primroot(v_1) \sim \primroot(v_2)$. So there exist
  $x,y \in\Sigma^*$ such that
  $v_1\in (xy)^*$ and $v_2\in (yx)^*$. By Lemma \ref{lem:outputform},
  we obtain that $v_1x_1v_2\in x(yx)^*$. Then it suffices to take
  $\ell = i$, $w = x_0$, $t_1 = x$, $t_2 = yx$ and $t_3 = x_2$ to conclude
    the proof.
    
  \item Otherwise, we have $v_1=\epsilon$. We decompose $x_1$ as
    $x_1=y_1y_2$ where $y_1=\out_1[j,n]$ and $y_2=\out_2[j,n]$. 

    We again distinguish two cases:
\begin{enumerate}
\item We first consider the case when $|y_1|=|\out_1[j,n]|> K$. In
  this case, we can as before decompose the input word $u[j..n]$ to
  identify a loop. More precisely, there exists a loop $(r,s)$ in
  $\rho$ such that $r\geq j$, $|\out_1[j,r]|< K$ and $1\leq
  |\out_1[r,s]|\leq K$. This loop gives a decomposition of $u_3$ as
  $u_4u_5u_6$ . We will then apply Property $\ppt$ to the two loops
  $(i,j)$ and $(r,s)$. The loop $(r,s)$ gives a decomposition of $y_1$
  as $z_0w_1z_1$, $y_2$ as $z_2w_2z_3$ and $x_4$ as $z_4w_3z_5$. By
  Property $\ppt$, there exist words $\beta_i$, $i\in\{1,\ldots,5\}$,
  and $c_1,c'_1,c_2,c'_2,f,g$ such that, for all $k_1,k_2\geq 0$,
    $$
    \begin{array}{c}
      f(k_1,k_2)x_0v_1^{\eta_1}z_0w_1^{\eta_2}z_1z_2w_2^{\eta_2}z_3v_2^{\eta_1}x_2\\
      x_3v_3^{\eta_1}z_4w_3^{\eta_2}z_5g(k_1,k_2)
      = \beta_1\beta_2^{k_1}\beta_3\beta_4^{k_2}\beta_5
    \end{array}
    $$
    where $\eta_i=k_ic_i+c'_i$, $i\in\{1,2\}$. Recall that
    $w_1\neq\varepsilon$ and $v_2\neq\varepsilon$. As a consequence,
    we can, using sufficiently large values of $k_1$ and $k_2$ and
    applying Lemma~\ref{lem:fundamental}, prove that
    $\primroot(w_1)\sim\primroot(\beta_4)$, that
    $\primroot(v_2)\sim\primroot(\beta_4)$, and thus deduce that
    $\primroot(w_1)\sim \primroot(v_2)$. Therefore there exist $x,y$
    such that $v_2 \in (yx)^*$ and $w_1\in (xy)^*$ from which we
    deduce that $w_1z_1z_2w_2z_3v_2\in x(yx)^*$. Recall that by the
    choice of the loop $(r,s)$ we have $|z_0|\leq K$. We can thus
    define $\ell = i$, $w = x_0$, $t_1 = z_0x$, $t_2 = yx$ and $t_3 =
    x_2$ to obtain the result.

  \item The last case is when $|y_1|=|\out_1[j,n]|\leq K$. We consider
    the length of $y_2=\out_2[j,n]$. First observe that if we have
    $|y_2|\leq K$ then we are done. Indeed, we can define $\ell=j$,
    $t_1=y_1$, $t_2=y_2$ and $t_3=v_2x_2$. It is routine to verify
    that the conditions of Property $\pptone$ are fulfilled.

    We thus suppose that $|y_2|>K$. In this case, we can as before
    identify a loop $(r,s)$ in the run $\rho$ such that $r\geq j$,
    $\out_2[s,n]\leq K$ and $1\leq \out_2[r,s]\leq K$. We do not give
    the details, but one can apply Property $\ppt$ to the two loops
    $(i,j)$ and $(r,s)$ and use the fact that
    $\out_2[i,j]\neq\varepsilon$ and $\out_2[r,s]\neq\varepsilon$ to
    prove that $\primroot(\out_2[i,j]) \sim
    \primroot(\out_2[r,s])$. Then, there exist $x,y$ such that
    $\out_2[r,s] \in (xy)^*$ and $\out_2[i,j] \in (yx)^*$ from which we
    deduce that $\out_2[i,s] \in (xy)^*x$.

    Finally, we let $\ell = i$, $w = x_0$, $t_1 =
    \out_1[i,n]\out_2[s,n]$, $t_2 = xy$ and $t_3 = xx_2$ to obtain
    the result. 
\end{enumerate}
\end{enumerate}
\end{IEEEproof}

%% file: forward.tex
\subsection{From \eZNFT to \NFT}

We state the following Lemma whose proof is similar to that of
Lemma~\ref{lem:simpleloopsruns}:
\begin{lemma}\label{lm:loop-eZNFT}
  Let $T\in \ZNFT$ with $m$ states. Let $o$ the maximal length of an
  output word in a transition of $T$ and $K=2.o.m^3.|\Sigma|$. Let
  $\rho$ be a run on a word $u$ of length $n$. We write $\rho$ as the
  sequence $(p_1,1)\dots (p_n,n)(q_{n-1},n-1)\dots (q_1,1)(r_2,2)\dots
  (r_{n+1},n+1)$ and let $q_n=p_n$ and $r_1=q_1$. Let two indices $1\leq i
  \leq j \leq n$. Then, we have:
\begin{enumerate}
\item if $|\out_3[i,j]| > K$, there
  exists a loop $(k_1,k_2)$ in $\rho$ with $i\leq k_1 < k_2\leq j$ such that
  \begin{enumerate}
    \item $|\out_3[i,k_1]|\leq K$
    \item $1\leq |\out_3[k_1,k_2]| \leq K$
  \end{enumerate}
\item if $|\out_3[i,j]| > K$, there
  exists a loop $(k_1,k_2)$ in $\rho$ with $i\leq k_1 < k_2\leq j$ such that
  \begin{enumerate}
    \item $|\out_3[k_2,j]|\leq K$
    \item $1\leq |\out_3[k_1,k_2]| \leq K$
  \end{enumerate}
\item if $|\out_1[i,j]| > K$, there
  exists a loop $(k_1,k_2)$ in $\rho$ with $i\leq k_1 < k_2\leq j$ such that
  \begin{enumerate}
    \item $|\out_1[i,k_1]|\leq K$
    \item $1\leq |\out_1[k_1,k_2]| \leq K$
  \end{enumerate}
\item if $|\out_1[i,j]| > K$, there
  exists a loop $(k_1,k_2)$ in $\rho$ with $i\leq k_1 < k_2\leq j$ such that
  \begin{enumerate}
    \item $|\out_1[k_2,j]|\leq K$
    \item $1\leq |\out_1[k_1,k_2]| \leq K$
  \end{enumerate}
\end{enumerate}
\end{lemma}

\smallskip\noindent\textbf{Proof of Proposition~\ref{prop:pptimpliesppttwo}}

\begin{IEEEproof}
  We let $T=(Q,q_0,F,\Delta)$ and $K=2.o.m^3.|\Sigma|$. Recall that as
  $T'\in\eZNFT$, we have $\out_2[1,n]=\epsilon$.

  Let us define the position $\ell$ as the largest positive integer
  less than or equal to $n$ such that $\out_1[\ell,n]=\epsilon$. 

  We first observe that if $|\out_3[1,\ell]|\leq K$, then we are
  done, by considering $\ell_1=\ell_2=\ell$. Indeed, we then consider
  $w=\out_1[1,\ell]$, $w' = \out_3[\ell,n+1]$,
  $t_1=\out_3[1,\ell]$, and $t_2=t_3=\epsilon$.

  Thus, we now suppose that we have $|\out_3[1,\ell]| > K$. In this
  case, we can apply Lemma~\ref{lm:loop-eZNFT}, case $1)$: there exists
  a loop $(k_1,k_2)$ such that $|\out_3[1,k_1]|
  \leq K$ and $1\leq |\out_3[k_1,k_2]| \leq K$.
  
  We again distinguish two cases:
    
\noindent\textit{Case I: } $|\out_3[k_2,\ell]|\leq K$. For this case, we
again distinguish three cases, depending on the value of
$\out_1[k_1,k_2]$ and on the length of $|\out_1[k_2,\ell]|\leq K$:
\begin{enumerate}
\item if we have $\out_1[k_1,k_2]\neq \epsilon$.  We will prove that
  the output word $\out_1[k_1,n]\out_3[1,\ell]$ has the
  expected form ($t_1t_2^*t_3$).  Therefore we use the $\ppt$-property
  on the loop $(k_1,k_2)$ with an additional empty loop. We define:
  $$
  \begin{array}{ll}
    w   &=\out_1[1,k_1]\\
    x_1 &=\out_1[k_1,k_2]\\
    y   &=\out_1[k_2,n]\out_3[1,k_1]\\
    x_2 &=\out_3[k_1,k_2]\\
    z   &=\out_3[k_2,\ell]\\
    w'  &=\out_3[\ell,n+1]
  \end{array}
  $$
  Property $\ppt$ entails that there exist
  $\beta_1,\beta_2,\beta_3,f,g,c,c'$ such that, for all $k\geq 0$,
  $$
  f(k)wx_1^{kc+c'}yx_2^{kc+c'}zw'g(k)=\beta_1 \beta_2^{k} \beta_3
  $$
  As we have $x_1\neq \epsilon$, and $x_2\neq\epsilon$, this entails,
  thanks to the fundamental lemma (Lemma~\ref{lem:fundamental}), that
  $\primroot(x_1)\sim \primroot(x_2)$. Let $t_2$ be
  $\primroot(x_1)$. We can write $t_2=z_1z_2$ and
  $\primroot(x_2)=z_2z_1$. As a consequence, we obtain that $x_1yx_2$
  is of the form $t_2^*.z_1$ by Lemma~\ref{lem:outputform}. We can
  thus set $t_1=\epsilon$, $t_3=z_1.z$, $\ell_1=k_1$ and
  $\ell_2=\ell$. It is routine to verify that words $w,w',t_1,t_2,t_3$
  verify the conditions of $\ppttwo$-property.

  This case is depicted on Figure~\ref{fig:case1}.


\input{figures/case1}

\item if we have $\out_1[k_1,k_2]= \epsilon$ and
  $|\out_1[k_2,\ell]|\leq K$. We will show that the result is easy.
  Indeed, consider $\ell_1=k_1$, $\ell_2=\ell$,
  $t_1=\out_1[k_1,n]\out_3[1,k_1]$, $t_2=\out_3[k_1,k_2]$, and
  $t_3=\out_3[k_2,\ell]$. It is routine to verify that all the
  requirements of $\ppttwo$-property are met.

  This case is depicted on Figure~\ref{fig:case2}.


\input{figures/case2}

\item last, if we have $\out_1[k_1,k_2]= \epsilon$ and
  $|\out_1[k_2,\ell]| > K$.  In this case, we will have to identify a
  loop in this part ($[k_2,\ell]$) of the input word, to prove the
  expected form of the output words. Formally, we apply
  Lemma~\ref{lm:loop-eZNFT} as we did before, except that we are
  interested in the output produced in the first pass of the \ZNFT,
  and not in that produced in the third pass. We thus apply case $3)$
  of Lemma~\ref{lm:loop-eZNFT}. We can thus exhibit a loop $(j_1,j_2)$
  with $k_2\leq j_1<j_2\leq \ell-1$ such that $|\out_1[k_2,j_1]| \leq
  K$ and $1\leq |\out_1[j_1,j_2]| \leq K$.
  
  We are now ready to prove that the output word
  $\out_1[k_1,n]\out_3[1,\ell]$ has the expected form
  ($t_1t_2^*t_3$). To this aim, we define:
  $$
  \begin{array}{llll}
    u_1 &=u[1,k_1-1]          &     w   &=\out_1[1,k_1]\\
    u_2 &=u[k_1,k_2-1]        &     t_1 &=\out_1[k_1,j_1]\\
    u_3 &=u[k_2,j_1-1]        &     x_1 &=\out_1[j_1,j_2]\\
    u_4 &=u[j_1,j_2-1]        &     y   &=\out_1[j_2,n]\out_3[1,k_1]\\
    u_5 &=u[j_2,\ell-1]       &     x_2 &=\out_3[k_1,k_2]\\
    u_6 &=u[\ell,n]           &     z_1 &=\out_3[k_2,j_1]\\
        &                     &     z_2 &=\out_3[j_1,j_2]\\
        &                     &     z_3 &=\out_3[j_2,\ell]\\
        &                     &     w'  &=\out_3[\ell,n+1]
  \end{array}
  $$
  As $(k_1,k_2)$ and $(j_1,j_2)$ are loops, we can apply Property
  $\ppt$.
%
  Using the fundamental lemma, we can deduce that $\primroot(x_1)\sim
  \primroot(x_2)$, using a reasoning similar to that of the proof of
  Proposition~\ref{prop:outputsnecessary}.  Thus, we can set
  $t_2=\primroot(x_1)$, and write $t_2=\alpha_1\alpha_2$ such that
  $\primroot(x_2)=\alpha_2\alpha_1$, from which we deduce $x_1yx_2\in
  t_2^*\alpha_1$ (Lemma~\ref{lem:outputform}). Finally, we let
  $z=z_1z_2z_3$, $t_3=\alpha_1z$, $\ell_1=k_1$ and $\ell_2=\ell$. The
  reader can verify that all the requirements of $\ppttwo$-property
  are met.

  This case is depicted on Figure~\ref{fig:case3}.


\input{figures/case3}

\end{enumerate}

\noindent\textit{Case II:} we have $|\out_3[k_2,\ell]| > K$. We
distinguish three cases, according to the length of the word
$\out_1[k_2,\ell]$, and to the value of $\out_1[k_1,k_2]$:
\begin{enumerate}
\item if we have $|\out_1[k_2,\ell]| \leq K$, we distinguish two cases:
\begin{enumerate}
\item We first consider the case when $\out_1[k_1,k_2]=\epsilon$. In
  this case, we can simply define $\ell_1=\ell_2=k_1$, and verify that
  the conditions of the $\ppttwo$-property are met.  This case is
  depicted on Figure~\ref{fig:case4}.

   
\input{figures/case4}

\item The second case is when $\out_1[k_1,k_2]\neq \epsilon$. This
  case is easy as we can show that $\primroot(\out_1[k_1,k_2]) \sim
  \primroot(\out_3[k_1,k_2])$, and deduce the expected form for the
  output words, by setting $\ell_2=k_2$.  This case is depicted on
  Figure~\ref{fig:case5}.

\input{figures/case5}
\end{enumerate}

\item if we have $|\out_1[k_2,\ell]| > K$ and $\out_1[k_1,k_2]\neq
  \epsilon$. As $|\out_1[k_2,\ell]| > K$, we can apply
  Lemma~\ref{lm:loop-eZNFT}, case $4)$, to identify a loop $(j_1,j_2)$
  such that $|\out_1[j_2,\ell]|\leq K$ and $1\leq
  |\out_1[j_1,j_2]|\leq K$. In this case, we set $\ell_1=k_1$ and
  $\ell_2=j_2$.

  There are three cases, according to $\out_3[j_1,j_2]$ and
  $\out_3[k_2,j_1]$:
  \begin{enumerate}
  \item We first consider the case when $\out_3[j_1,j_2]\neq
    \epsilon$. In this case, using Property $\ppt$, we can show that
    $\primroot(\out_1[k_1,k_2])\sim \primroot(\out_1[j_1,j_2])\sim
    \primroot(\out_3[k_1,k_2])\sim \primroot(\out_3[j_1,j_2])$. This
    allows to prove the expected form of the output words.
  
  \item Second, we suppose that $\out_3[j_1,j_2] = \epsilon$ and that
    $\out_3[k_2,j_1]\leq K$. In this case, we can use the word $t_3$
    to cover the output word $\out_3[k_2,j_1]$. Last, using a
    reasoning on word combinatorics, we can prove that
    $\primroot(\out_1[k_1,k_2])\sim \primroot(\out_1[j_1,j_2])\sim
    \primroot(\out_3[k_1,k_2])$ and conclude.
  
    Cases a) and b) are depicted on Figure~\ref{fig:case6}.

\input{figures/case6}

\item Last, we consider the case $\out_3[j_1,j_2] = \epsilon$ and
  $\out_3[k_2,j_1]> K$. By Lemma~\ref{lm:loop-eZNFT}, case $2)$, there
  exists a loop $(p_1,p_2)$ included in the interval $[k_2,j_1]$ such
  that $|\out_3[p_2,j_1]|\leq K$ and $1\leq |\out_3[p_1,p_2]|\leq
  K$. We claim that the result holds.  The only difficult property is
  the fact the output word has the expected form ($t_1t_2^*t_3$). This
  can be proven using word combinatorics, by showing, using the
  Property $\ppt$, that $\primroot(\out_1[k_1,k_2])\sim
  \primroot(\out_1[j_1,j_2])\sim
  \primroot(\out_3[k_1,k_2])\sim\primroot(\out_3[p_1,p_2])$.  
\end{enumerate}

\item last, if we have $|\out_1[k_2,\ell]| > K$ and
  $\out_1[k_1,k_2]=\epsilon$. We first let $\ell_1=k_1$. As we have
  $|\out_1[k_2,\ell]| > K$, we can apply Lemma~\ref{lm:loop-eZNFT},
  case $3)$, to identify a loop $(j_1,j_2)$ included in the interval
  $(k_2,\ell)$ such that $|\out_1[k_2,j_1]|\leq K$ and $1\leq
  |\out_1[j_1,j_2]|\leq K$. We distinguish two cases:
  \begin{enumerate}
  \item if $|\out_1[j_2,\ell]|\leq K$. We define $\ell_2=j_2$. We
    consider the value of $\out_3[j_1,j_2]$.

    If we have $\out_3[j_1,j_2]\neq \epsilon$, then we can
    conclude. Indeed, using word combinatorics, we can prove
    $\primroot(\out_1[j_1,j_2])\sim\primroot(\out_3[k_1,k_2])\sim\primroot(\out_3[j_1,j_2])$
    and prove that the output word $\out_1[k_1,n]\out_3[1,j_2]$ has
    the expected form. This case is depicted on Figure~\ref{fig:case8}.

\input{figures/case8}

    Otherwise, we have $\out_3[j_1,j_2]= \epsilon$. For this case we
    distinguish two cases:
    \begin{enumerate}
    \item if $\out_3[k_2,j_1]\leq K$: we can conclude
      directly. Indeed, it is easy to show that
      $\primroot(\out_1[j_1,j_2])\sim\primroot(\out_3[k_1,k_2])$. The word
      $\out_3[k_2,j_2]$ is not necessarily conjugated with the
      previous words, but its length is less than $K$ by hypothesis,
      thus we can use the word $t_3$ to handle this part of the output.
      This case is depicted on Figure~\ref{fig:case9}.

\input{figures/case9}

    \item if $\out_3[k_2,j_1] > K$: we will apply
      Lemma~\ref{lm:loop-eZNFT}, case $2)$, to identify a loop
      $(p_1,p_2)$ included in the interval $(k_2,j_1)$ such that
      $|\out_3[p_2,j_1]|\leq K$ and $1\leq
      |\out_3[p_1,p_2]|\leq K$. Then we can prove that
      $\primroot(\out_1[j_1,j_2])\sim\primroot(\out_3[k_1,k_2])\sim\primroot(\out_3[p_1,p_2])$
      and conclude.

    \end{enumerate}

  \item if $|\out_1[j_2,\ell]|> K$. We can apply
    Lemma~\ref{lm:loop-eZNFT}, case $4)$, to identify a loop
    $(p_1,p_2)$ included in the interval $(j_2,\ell)$ such that
    $|\out_1[p_2,\ell]|\leq K$ and $1\leq
    |\out_1[p_1,p_2]|\leq K$. In the sequel, we let $\ell_2$ be
    $p_2$ and $\ell_1$ be $k_1$. We let $\alpha=\out_3[p_1,p_2]$ and
    $\alpha'=\out_3[j_1,j_2]$. 
The situation is depicted on Figure~\ref{fig:case11}.

\input{figures/case11}

    We distinguish five cases:
    \begin{enumerate}
    \item if $\alpha\neq \epsilon$, we conclude easily by showing that
      $\primroot(\out_1[j_1,j_2])\sim\primroot(\out_1[p_1,p_2])
      \sim\primroot(\out_3[k_1,k_2])\sim\primroot(\out_3[p_1,p_2])$.
    \item if $\alpha= \epsilon$ and $|\out_3[j_2,p_1]|>K$: we can
      identify a loop in $\rho$, included in the interval
      $[j_2,p_1]$, such that $\out_3$ is non-empty on this
      loop. We can then derive the result.
    \item if $\alpha= \epsilon$, $|\out_3[j_2,p_1]|\leq K$ and
      $\alpha'\neq \epsilon$, then we can show that
      $\primroot(\out_1[j_1,j_2])\sim\primroot(\out_1[p_1,p_2])
      \sim\primroot(\out_3[k_1,k_2])\sim\primroot(\out_3[j_1,j_2])$, and
      conclude as the output $\out_3[j_2,p_2]$ has length less
      than $K$ ($t_3$ can be defined so as to cover these words).
    \item if $\alpha= \epsilon$, $|\out_3[j_2,p_1]|\leq K$,
      $\alpha'= \epsilon$ and $|\out_3[k_2,j_1]|>K$, we can identify
      a loop inside the interval $[k_2,j_1]$. This loop can be used
      to prove the result, as we know that the length of the word
      $\out_3[j_1,p_2]$ is less than $K$.
    \item else, \emph{i.e.} if $\alpha= \epsilon$,
      $|\out_3[j_2,p_1]|\leq K$, $\alpha'= \epsilon$ and
      $|\out_3[k_2,j_1]|\leq K$, then we are done as $t_3$ can be
      defined as $\out_3[k_2,p_2]$.
    \end{enumerate}

  \end{enumerate}

\end{enumerate}         

\end{IEEEproof}

\smallskip\noindent\textbf{Construction of $T''$ from $T'$}

We provide here some additional details for the definition of the \NFT $T''$
from the \eZNFT $T'$.

First, the transducer $T''$ should, in a single forward pass, simulate
the three passes (forward, backward, and forward) of $T'$. Therefore
it maintains a triple of states of $T'$ and the current symbol.

Second, it uses three modes: before the guess of position $\ell_1$,
between positions $\ell_1$ and $\ell_2$, and after position $\ell_2$.

Third, it should guess the words of bounded length $t_1$, $t_2$ and
$t_3$, and two additional words $x$ and $y$ of bounded length ($\leq
3.K$) which intuitively correspond to words $\out_3[1,\ell_1]$ and
$\out_1[\ell_2,n]$ (see property $\ppttwo$). 

Last, it verifies in the different modes that the output has the
expected form, and produces in a forward manner the overall output
word. Therefore it distinguishes between different cases, whether $t_1$
is a prefix of $\out_1[\ell_1,\ell_2]$ or whether $t_1$ also covers
$\out_1[\ell_2,n]$ or $\out_3[1,\ell_1]$, or even
$\out_3[\ell_1,\ell_2]$. It manipulates pointers in the different words
of bounded length it has guessed to verify the form of the output, and
to produce the correct output, as we did in the construction of $T'$.

%% file: figures/case1.tex
\usetikzlibrary{decorations.pathreplacing}

\begin{figure}[t]
   \centering
   \scriptsize
   \begin{tikzpicture}[]

     \draw [->,>=latex] 
       (1,0)    -- (7,0) ;
         \draw [->,>=latex] 
   (1,1.5)    -- (7,1.5) ;

     \draw[densely dotted] (1,-.5) -- (1,1.7);
     \draw[densely dotted] (2.5,-.5) -- (2.5,1.7);
     \draw[densely dotted] (4,-.5) -- (4,1.7);
     \draw[densely dotted] (5.5,-.5) -- (5.5,1.7);
     \draw[densely dotted] (7,-.5) -- (7,1.7);

      \draw (0,0) node {$\out_1$};
      \draw (0,1.5) node {$\out_3$};

     \draw (1,-.7) node {$1$};
     \draw (2.5,-.7) node {$k_1=\ell_1$};
     \draw (4,-.7) node {$k_2$};
     \draw (5.5,-.7) node {$\ell=\ell_2$};
     \draw (7,-.7) node {$n$};

    \draw [decorate,decoration={brace,amplitude=5pt}]
(1.1,1.65) -- (2.4,1.65) node [black,midway,yshift=9pt] {$\le K$};
         \draw (3.2,1.25) node {$x_2\neq\epsilon$};
         \draw (4.7,1.25) node {$z$};
         \draw (6.2,1.25) node {$w'$};
\draw [decorate,decoration={brace,amplitude=5pt}]
(4.1,1.65) -- (6.9,1.65) node [black,midway,yshift=9pt] {$\le K$};

         \draw (1.7,.15) node {$w$};
         \draw (3.2,.15) node {$x_1\neq \epsilon$};
         \draw (6.2,.15) node {$\epsilon$};

      \end{tikzpicture}
\caption{ \label{fig:case1} Decomposition of the output for case I.1)}
\end{figure}
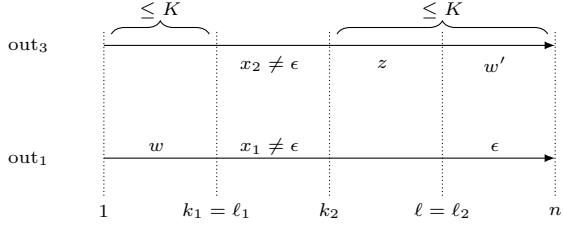

%% file: figures/case2.tex
\usetikzlibrary{decorations.pathreplacing}

\begin{figure}[t]
   \centering
   \scriptsize
   \begin{tikzpicture}[]

     \draw [->,>=latex] 
       (1,0)    -- (7,0) ;
         \draw [->,>=latex] 
   (1,1.5)    -- (7,1.5) ;

     \draw[densely dotted] (1,-.5) -- (1,1.7);
     \draw[densely dotted] (2.5,-.5) -- (2.5,1.7);
     \draw[densely dotted] (4,-.5) -- (4,1.7);
     \draw[densely dotted] (5.5,-.5) -- (5.5,1.7);
     \draw[densely dotted] (7,-.5) -- (7,1.7);

      \draw (0,0) node {$\out_1$};
      \draw (0,1.5) node {$\out_3$};

     \draw (1,-.7) node {$1$};
     \draw (2.5,-.7) node {$k_1=\ell_1$};
     \draw (4,-.7) node {$k_2$};
     \draw (5.5,-.7) node {$\ell=\ell_2$};
     \draw (7,-.7) node {$n$};

    \draw [decorate,decoration={brace,amplitude=5pt}]
(1.1,1.65) -- (2.4,1.65) node [black,midway,yshift=9pt] {$\le K$};
         \draw (1.7,1.25) node {$b$};
         \draw (3.2,1.25) node {$t_2\neq\epsilon$};
         \draw (4.7,1.25) node {$t_3$};
\draw [decorate,decoration={brace,amplitude=5pt}]
(4.1,1.65) -- (6.9,1.65) node [black,midway,yshift=9pt] {$\le K$};

         \draw (3.2,.15) node {$\epsilon$};
         \draw (4.7,.15) node {$a$};
         \draw [decorate,decoration={brace,amplitude=5pt}]
(5.4,-.15) -- (4.1,-.15)  node [black,midway,yshift=-9pt] {$\le K$};

         \draw (6.2,.15) node {$\epsilon$};

         \draw (7.5,.75) node {$t_1 = ab$};

%
%



   \end{tikzpicture}
\caption{ \label{fig:case2} Decomposition of the output for case I.2)}
\end{figure}
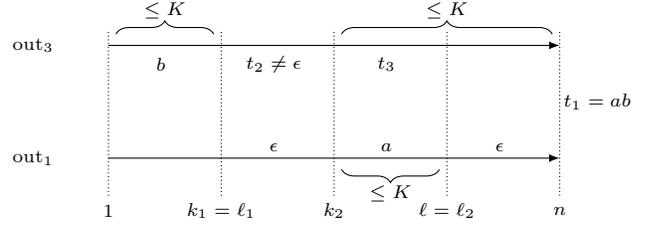

%% file: figures/case3.tex
\usetikzlibrary{decorations.pathreplacing}

\begin{figure}[t]
   \centering
   \scriptsize
   \begin{tikzpicture}[]

     \draw [->,>=latex] 
       (0.5,0)    -- (8.5,0) ;
         \draw [->,>=latex] 
   (0.5,1.5)    -- (8.5,1.5) ;

     \draw[densely dotted] (0.5,-.5) -- (0.5,1.7);
     \draw[densely dotted] (1.5,-.5) -- (1.5,1.7);
     \draw[densely dotted] (3,-.5) -- (3,1.7);
     \draw[densely dotted] (4.5,-.5) -- (4.5,1.7);
     \draw[densely dotted] (6,-.5) -- (6,1.7);
     \draw[densely dotted] (7.5,-.5) -- (7.5,1.7);
     \draw[densely dotted] (8.5,-.5) -- (8.5,1.7);

      \draw (0,0) node {$\out_1$};
      \draw (0,1.5) node {$\out_3$};

     \draw (.5,-.7) node {$1$};
     \draw (1.5,-.7) node {$k_1=\ell_1$};
     \draw (3,-.7) node {$k_2$};
     \draw (4.5,-.7) node {$j_1$};
     \draw (6,-.7) node {$j_2$};
     \draw (7.5,-.7) node {$\ell=\ell_2$};
     \draw (8.5,-.7) node {$n$};

        \draw [decorate,decoration={brace,amplitude=5pt}]
(7.4,1.15) -- (2.4,1.15) node [black,midway,yshift=-9pt] {$t_3$};

    \draw [decorate,decoration={brace,amplitude=5pt}]
(.6,1.65) -- (1.4,1.65) node [black,midway,yshift=9pt] {$\le K$};
         \draw (2.2,1.25) node {$x_2\neq \epsilon$};
         \draw (3.7,1.25) node {$z_1$};
         \draw (5.2,1.25) node {$z_2$};
         \draw (6.7,1.25) node {$z_3$};
         \draw (8,1.25) node {$w'$};
\draw [decorate,decoration={brace,amplitude=5pt}]
(3.1,1.65) -- (8.4,1.65) node [black,midway,yshift=9pt] {$\le K$};

\draw [decorate,decoration={brace,amplitude=5pt}]
(1.6,.15) -- (4.4,.15) node [black,midway,yshift=9pt] {$t_1$};

         \draw (1,.15) node {$w$};
         \draw (2.2,.15) node {$\epsilon$};
         \draw (5.2,.15) node {$x_1\neq\epsilon$};
         \draw (8,.15) node {$\epsilon$};
         \draw [decorate,decoration={brace,amplitude=5pt}]
(4.4,-.15) -- (3.1,-.15)  node [black,midway,yshift=-9pt] {$\le K$};

%
%



   \end{tikzpicture}
\caption{ \label{fig:case3} Decomposition of the output for case I.3)}
\end{figure}
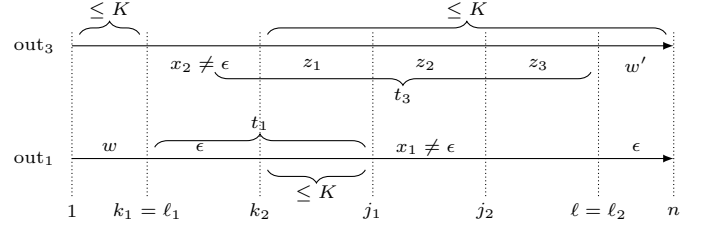

%% file: figures/case4.tex
\usetikzlibrary{decorations.pathreplacing}

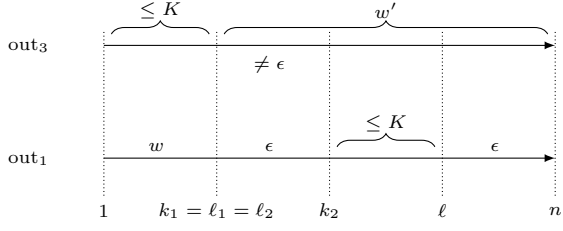
\begin{figure}[t]
   \centering
   \scriptsize
   \begin{tikzpicture}[]

     \draw [->,>=latex] 
       (1,0)    -- (7,0) ;
         \draw [->,>=latex] 
   (1,1.5)    -- (7,1.5) ;

     \draw[densely dotted] (1,-.5) -- (1,1.7);
     \draw[densely dotted] (2.5,-.5) -- (2.5,1.7);
     \draw[densely dotted] (4,-.5) -- (4,1.7);
     \draw[densely dotted] (5.5,-.5) -- (5.5,1.7);
     \draw[densely dotted] (7,-.5) -- (7,1.7);

      \draw (0,0) node {$\out_1$};
      \draw (0,1.5) node {$\out_3$};

     \draw (1,-.7) node {$1$};
     \draw (2.5,-.7) node {$k_1=\ell_1=\ell_2$};
     \draw (4,-.7) node {$k_2$};
     \draw (5.5,-.7) node {$\ell$};
     \draw (7,-.7) node {$n$};

    \draw [decorate,decoration={brace,amplitude=5pt}]
(1.1,1.65) -- (2.4,1.65) node [black,midway,yshift=9pt] {$\le K$};
         \draw (3.2,1.25) node {$\neq\epsilon$};
\draw [decorate,decoration={brace,amplitude=5pt}]
(2.6,1.65) -- (6.9,1.65) node [black,midway,yshift=9pt] {$w'$};

         \draw (1.7,.15) node {$w$};
         \draw (3.2,.15) node {$\epsilon$};
         \draw (6.2,.15) node {$\epsilon$};
         \draw [decorate,decoration={brace,amplitude=5pt}]
   (4.1,.15) -- (5.4,.15)  node [black,midway,yshift=9pt] {$\le K$};

%
%



   \end{tikzpicture}
\caption{ \label{fig:case4} Decomposition of the output, case II.1).a)}
\end{figure}

%% file: figures/case5.tex
\usetikzlibrary{decorations.pathreplacing}

\begin{figure}[t]
   \centering
   \scriptsize
   \begin{tikzpicture}[]

     \draw [->,>=latex] 
       (1,0)    -- (7,0) ;
         \draw [->,>=latex] 
   (1,1.5)    -- (7,1.5) ;

     \draw[densely dotted] (1,-.5) -- (1,1.7);
     \draw[densely dotted] (2.5,-.5) -- (2.5,1.7);
     \draw[densely dotted] (4,-.5) -- (4,1.7);
     \draw[densely dotted] (5.5,-.5) -- (5.5,1.7);
     \draw[densely dotted] (7,-.5) -- (7,1.7);

      \draw (0,0) node {$\out_1$};
      \draw (0,1.5) node {$\out_3$};

     \draw (1,-.7) node {$1$};
     \draw (2.5,-.7) node {$k_1=\ell_1$};
     \draw (4,-.7) node {$k_2=\ell_2$};
     \draw (5.5,-.7) node {$\ell$};
     \draw (7,-.7) node {$n$};

    \draw [decorate,decoration={brace,amplitude=5pt}]
(1.1,1.65) -- (2.4,1.65) node [black,midway,yshift=9pt] {$\le K$};
         \draw (3.2,1.25) node {$\neq\epsilon$};
\draw [decorate,decoration={brace,amplitude=5pt}]
(4.1,1.65) -- (6.9,1.65) node [black,midway,yshift=9pt] {$w'$};

         \draw (1.7,.15) node {$w$};
         \draw (3.2,.15) node {$\neq\epsilon$};
         \draw [decorate,decoration={brace,amplitude=5pt}]
   (4.1,.15) -- (5.4,.15)  node [black,midway,yshift=9pt] {$\le K$};

         \draw (6.2,.15) node {$\epsilon$};

%
%



   \end{tikzpicture}
\caption{ \label{fig:case5} Decomposition of the output for case II.1).b)}
\end{figure}
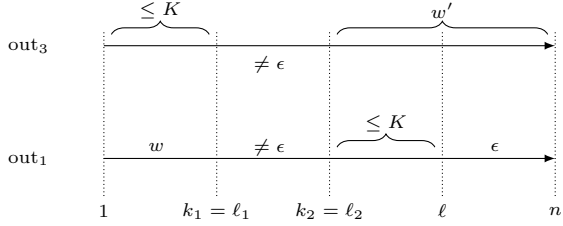

%% file: figures/case6.tex
\usetikzlibrary{decorations.pathreplacing}

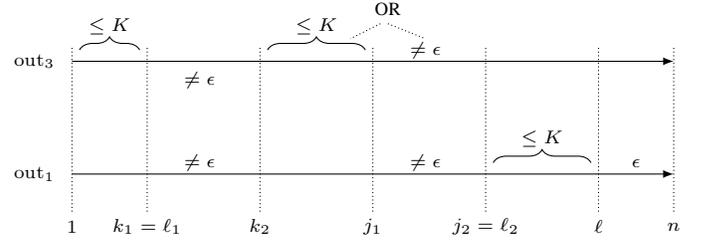
\begin{figure}[t]
   \centering
   \scriptsize
   \begin{tikzpicture}[]

     \draw [->,>=latex] 
       (0.5,0)    -- (8.5,0) ;
         \draw [->,>=latex] 
   (0.5,1.5)    -- (8.5,1.5) ;

     \draw[densely dotted] (0.5,-.5) -- (0.5,1.7);
     \draw[densely dotted] (1.5,-.5) -- (1.5,1.7);
     \draw[densely dotted] (3,-.5) -- (3,1.7);
     \draw[densely dotted] (4.5,-.5) -- (4.5,1.7);
     \draw[densely dotted] (6,-.5) -- (6,1.7);
     \draw[densely dotted] (7.5,-.5) -- (7.5,1.7);
     \draw[densely dotted] (8.5,-.5) -- (8.5,1.7);

      \draw (0,0) node {$\out_1$};
      \draw (0,1.5) node {$\out_3$};

     \draw (.5,-.7) node {$1$};
     \draw (1.5,-.7) node {$k_1=\ell_1$};
     \draw (3,-.7) node {$k_2$};
     \draw (4.5,-.7) node {$j_1$};
     \draw (6,-.7) node {$j_2=\ell_2$};
     \draw (7.5,-.7) node {$\ell$};
     \draw (8.5,-.7) node {$n$};

    \draw [decorate,decoration={brace,amplitude=5pt}]
(.6,1.65) -- (1.4,1.65) node [black,midway,yshift=9pt] {$\le K$};
         \draw (2.2,1.25) node {$\neq \epsilon$};
\draw [decorate,decoration={brace,amplitude=5pt}]
(3.1,1.65) -- (4.4,1.65) node [black,midway,yshift=9pt] {$\le K$};
         \draw (5.2,1.65) node {$\neq \epsilon$};

         \draw (2.2,.15) node {$\neq\epsilon$};
         \draw (5.2,.15) node {$\neq\epsilon$};
         \draw (8,.15) node {$\epsilon$};
         \draw [decorate,decoration={brace,amplitude=5pt}]
(6.1,.15) -- (7.4,.15)  node [black,midway,yshift=9pt] {$\le K$};

     \draw (4.7,2.2) node {OR};
     \draw[densely dotted] (4.6,2)  -- (4.2,1.8);
     \draw[densely dotted] (4.7,2) -- (5.2,1.8);

%
%



   \end{tikzpicture}
\caption{ \label{fig:case6} Decomposition of the output for case II.2).a) and b)}
\end{figure}

%% file: figures/case8.tex
\usetikzlibrary{decorations.pathreplacing}

\begin{figure}[t]
   \centering
   \scriptsize
   \begin{tikzpicture}[]

     \draw [->,>=latex] 
       (0.5,0)    -- (8.5,0) ;
         \draw [->,>=latex] 
   (0.5,1.5)    -- (8.5,1.5) ;

     \draw[densely dotted] (0.5,-.5) -- (0.5,1.7);
     \draw[densely dotted] (1.5,-.5) -- (1.5,1.7);
     \draw[densely dotted] (3,-.5) -- (3,1.7);
     \draw[densely dotted] (4.5,-.5) -- (4.5,1.7);
     \draw[densely dotted] (6,-.5) -- (6,1.7);
     \draw[densely dotted] (7.5,-.5) -- (7.5,1.7);
     \draw[densely dotted] (8.5,-.5) -- (8.5,1.7);

      \draw (0,0) node {$\out_1$};
      \draw (0,1.5) node {$\out_3$};

     \draw (.5,-.7) node {$1$};
     \draw (1.5,-.7) node {$k_1=\ell_1$};
     \draw (3,-.7) node {$k_2$};
     \draw (4.5,-.7) node {$j_1$};
     \draw (6,-.7) node {$j_2=\ell_2$};
     \draw (7.5,-.7) node {$\ell$};
     \draw (8.5,-.7) node {$n$};

    \draw [decorate,decoration={brace,amplitude=5pt}]
(.6,1.65) -- (1.4,1.65) node [black,midway,yshift=9pt] {$\le K$};
    \draw [decorate,decoration={brace,amplitude=5pt}]
(5.9,1.35) -- (.6,1.35) ;
         \draw (2.2,1.65) node {$\neq \epsilon$};
         \draw (5.2,1.65) node {$\neq \epsilon$};

         \draw (1.2,.15) node {$w$};
         \draw (2.2,.15) node {$\epsilon$};
                  \draw [decorate,decoration={brace,amplitude=5pt}]
(4.4,-.15)   -- (3.1,-.15) node [black,midway,yshift=-9pt] {$\le K$};

         \draw (5.2,.15) node {$\neq\epsilon$};
         \draw (8,.15) node {$\epsilon$};
         
         \draw [decorate,decoration={brace,amplitude=5pt}]
(1.6,.15) -- (4.4,.15) node [black,midway,yshift=9pt] {$t_1$};

         \draw [decorate,decoration={brace,amplitude=5pt}]
(7.4,-.15) -- (6.1,-.15)  node [black,midway,yshift=-9pt] {$\le K$};

         \draw [decorate,decoration={brace,amplitude=5pt}]
(4.6,.2) -- (8.4,.2)  ;

         \draw (4.8,.7) node {conjugate};
     \draw[densely dotted] (4.2,.7) -- (3.3,1.1);
     \draw[densely dotted] (5.4,.7) -- (6.5,.4);

   \end{tikzpicture}
   \caption{ \label{fig:case8} Decomposition of the output for case
     II.3).a), $\out_3[j_1,j_2]\neq\epsilon$}
\end{figure}
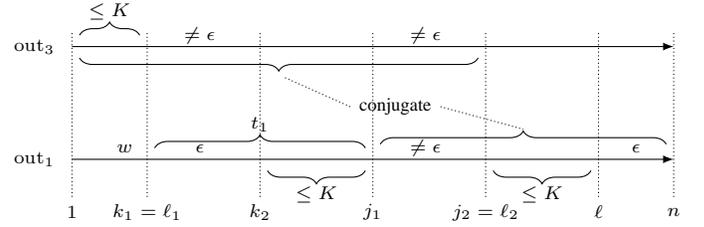

%% file: figures/case9.tex
\usetikzlibrary{decorations.pathreplacing}

\begin{figure}[t]
   \centering
   \scriptsize
   \begin{tikzpicture}[]

     \draw [->,>=latex] 
       (0.5,0)    -- (8.5,0) ;
         \draw [->,>=latex] 
   (0.5,1.5)    -- (8.5,1.5) ;

     \draw[densely dotted] (0.5,-.5) -- (0.5,1.7);
     \draw[densely dotted] (1.5,-.5) -- (1.5,1.7);
     \draw[densely dotted] (3,-.5) -- (3,1.7);
     \draw[densely dotted] (4.5,-.5) -- (4.5,1.7);
     \draw[densely dotted] (6,-.5) -- (6,1.7);
     \draw[densely dotted] (7.5,-.5) -- (7.5,1.7);
     \draw[densely dotted] (8.5,-.5) -- (8.5,1.7);

      \draw (0,0) node {$\out_1$};
      \draw (0,1.5) node {$\out_3$};

     \draw (.5,-.7) node {$1$};
     \draw (1.5,-.7) node {$k_1=\ell_1$};
     \draw (3,-.7) node {$k_2$};
     \draw (4.5,-.7) node {$j_1$};
     \draw (6,-.7) node {$j_2=\ell_2$};
     \draw (7.5,-.7) node {$\ell$};
     \draw (8.5,-.7) node {$n$};

    \draw [decorate,decoration={brace,amplitude=5pt}]
(.6,1.65) -- (1.4,1.65) node [black,midway,yshift=9pt] {$\le K$};
    \draw [decorate,decoration={brace,amplitude=5pt}]
(3.1,1.65) -- (4.4,1.65) node [black,midway,yshift=9pt] {$\le K$};
    \draw [decorate,decoration={brace,amplitude=5pt}]
(8.4,1.35) -- (6.1,1.35) node [black,midway,yshift=-9pt] {$w'$};
    \draw [decorate,decoration={brace,amplitude=5pt}]
(5.9,1.35) -- (2.5,1.35) node [black,midway,yshift=-9pt] {$t_3$};

         \draw (2.2,1.65) node {$\neq \epsilon$};
         \draw (5.2,1.65) node {$ \epsilon$};

         \draw (1,.15) node {$w$};
         \draw (2.2,.15) node {$\epsilon$};
                  \draw [decorate,decoration={brace,amplitude=5pt}]
(4.4,-.15)   -- (3.1,-.15) node [black,midway,yshift=-9pt] {$\le K$};

         \draw (5.2,.15) node {$\neq\epsilon$};
         \draw (8,.15) node {$\epsilon$};
         
         \draw [decorate,decoration={brace,amplitude=5pt}]
(1.6,.15) -- (4.4,.15) node [black,midway,yshift=9pt] {$t_1$};

         \draw [decorate,decoration={brace,amplitude=5pt}]
(7.4,-.15) -- (6.1,-.15)  node [black,midway,yshift=-9pt] {$\le K$};

   \end{tikzpicture}
\caption{ \label{fig:case9} Decomposition of the output for case II.3).a).i)}
\end{figure}
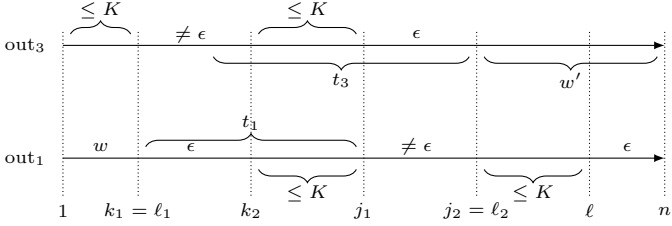

%% file: figures/case11.tex
\usetikzlibrary{decorations.pathreplacing}

\begin{figure*}[t]
   \centering
   \scriptsize
   \begin{tikzpicture}[]

     \draw [->,>=latex] 
       (0.5,0)    -- (10.5,0) ;
         \draw [->,>=latex] 
   (0.5,1.5)    -- (10.5,1.5) ;

     \draw[densely dotted] (0.5,-.5) -- (0.5,1.7); 
     \draw[densely dotted] (1.5,-.5) -- (1.5,1.7); 
     \draw[densely dotted] (2.5,-.5) -- (2.5,1.7); 
     \draw[densely dotted] (4.5,-.5) -- (4.5,1.7); 
     \draw[densely dotted] (5.5,-.5) -- (5.5,1.7); 
     \draw[densely dotted] (7,-.5) -- (7,1.7);     
     \draw[densely dotted] (8,-.5) -- (8,1.7);     
     \draw[densely dotted] (9.5,-.5) -- (9.5,1.7); 
     \draw[densely dotted] (10.5,-.5) -- (10.5,1.7); 

      \draw (0,0) node {$\out_1$};
      \draw (0,1.5) node {$\out_3$};

     \draw (.5,-.7) node {$1$};
     \draw (1.5,-.7) node {$k_1=\ell_1$};
     \draw (2.5,-.7) node {$k_2$};
     \draw (4.5,-.7) node {$j_1$};
     \draw (5.5,-.7) node {$j_2$};
     \draw (7,-.7) node {$p_1$};
     \draw (8,-.7) node {$p_2 = \ell_2$};
     \draw (9.5,-.7) node {$\ell$};
     \draw (10.5,-.7) node {$n$};

    \draw [decorate,decoration={brace,amplitude=5pt}]
(.6,1.65) -- (1.4,1.65) node [black,midway,yshift=9pt] {$\le K$};
    \draw [decorate,decoration={brace,amplitude=5pt}]
(4.6,1.65) -- (5.4,1.65) node [black,midway,yshift=9pt] {$\alpha'$};
    \draw [decorate,decoration={brace,amplitude=5pt}]
(7.1,1.65) -- (7.9,1.65) node [black,midway,yshift=9pt] {$\alpha$};


         \draw (2.2,1.65) node {$\neq \epsilon$};

         \draw (1,.15) node {$w$};
         \draw (2.2,.15) node {$\epsilon$};
                  \draw [decorate,decoration={brace,amplitude=5pt}]
(9.4,-.15)   -- (8.1,-.15) node [black,midway,yshift=-9pt] {$\le K$};

         \draw (5,.15) node {$\neq\epsilon$};
         \draw (7.5,.15) node {$\neq\epsilon$};
         \draw (10,.15) node {$\epsilon$};
         

         \draw [decorate,decoration={brace,amplitude=5pt}]
(4.4,-.15) -- (2.6,-.15)  node [black,midway,yshift=-9pt] {$\le K$};

   \end{tikzpicture}
\caption{ \label{fig:case11} Decomposition of the output for case II.3).b)}
\end{figure*}
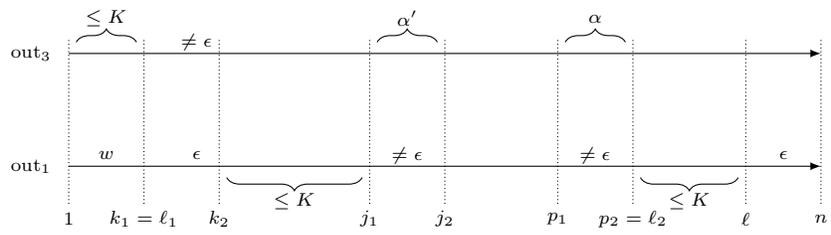